\documentclass[a4paper,11pt]{article}

\usepackage{mathptmx}       
\usepackage{helvet}         
\usepackage{courier}        
\usepackage[margin=1.14in]{geometry}
\usepackage{makeidx}         
\usepackage{graphicx}        
\usepackage{multicol}        
\usepackage[bottom]{footmisc}

\usepackage{amssymb}
\usepackage{cite}

\usepackage[T1]{fontenc}
\usepackage[utf8]{inputenc}
\usepackage{authblk}

\makeindex             

\begin{document}

\title{\huge \bf Bright solitary matter waves: formation, stability and interactions}
\author[1]{T. P. Billam}
\author[2]{A. L. Marchant}
\author[2]{S. L. Cornish}
\author[2]{S. A. Gardiner}
\author[3]{N. G. Parker\thanks{nick.parker@ncl.ac.uk}}
\affil[1]{\normalsize Jack Dodd Centre for Quantum Technology, Department of Physics, University of Otago, Dunedin, New Zealand}
\affil[2]{\normalsize Joint Quantum Centre (JQC) Durham-Newcastle, Department of Physics, Durham University, Durham, UK}
\affil[3]{\normalsize Joint Quantum Centre (JQC) Durham-Newcastle, School of Mathematics and Statistics, Newcastle University, Newcastle upon Tyne, UK}

%
%
\date{}
\maketitle

\begin{abstract}
In recent years, bright soliton-like structures composed of
gaseous Bose-Einstein condensates have been generated at ultracold temperature. The experimental capacity to precisely engineer the nonlinearity
and potential landscape experienced by these solitary waves offers an attractive
platform for fundamental study of solitonic structures.   The presence of three spatial dimensions and trapping implies that these are strictly distinct objects to the true soliton solutions.
Working within the zero-temperature mean-field description, we explore the solutions and stability of bright solitary waves, as well as their interactions.
Emphasis is placed on elucidating their similarities and differences to the true bright soliton.  The rich behaviour introduced in the bright solitary waves includes the collapse
instability and symmetry-breaking collisions.
We review the experimental formation and observation of bright solitary
matter waves to date, and compare to theoretical predictions.  Finally we discuss the current state-of-the-art of this area, including beyond-mean-field descriptions, exotic bright solitary waves, and proposals to exploit bright solitary waves in interferometry and as surface probes.
\end{abstract}


\section{Introduction}
\label{s_introduction}

\subsection{Gaseous Bose-Einstein condensates}
In 1925 Einstein predicted that an ideal and uniform gas of bosons, under conditions of sufficiently high density and/or low temperature, would begin to ``condense" into the single particle quantum state of zero energy \cite{einstein_1925}.  This phenomenon of Bose-Einstein condensation is now known to extend beyond gases into liquids and solids, being the underlying mechanism responsible for superfluidity in Helium and superconductivity \cite{tilley}.  But it is the gaseous form of this phenomenon that offers the purest and most controllable realization of this state of matter \cite{pethick_smith_2002,emergent_book,pitaevskii_stringari_2003}.

Since 1995, gaseous atomic Bose-Einstein condensates (BECs) have been generated in laboratories world-wide.  These gases are extremely dilute, with typical number densities of $10^{18}$--$10^{21}$ m$^{-3}$, and the onset of Bose-Einstein condensation occurs at ultracold temperatures of around $100$ nK \cite{pethick_smith_2002}.  Typically, experiments are run sufficiently far below this critical temperature that practically all the atoms enter the Bose-Einstein condensate, and the remaining thermal gas component becomes negligible.  Within the Bose-Einstein condensate, the individual de Broglie wavelengths of the atoms overlap, forming a single coherent matter wave that extends across the system.  From a theoretical perspective, this enables the description of the many-body system via a single mean-field wave equation.

Although these gases are dilute, the atom-atom interactions play a significant role and introduce a nonlinearity into the system.  At such low temperatures and densities, the interactions arise predominantly via elastic {\it s}-wave collisions, which are short-range and introduce a local cubic nonlinearity into the mean-field wave equation.  Furthermore, these interactions are usually repulsive.

The gases are formed and held within confining potentials produced via magnetic or optical fields.  These make the condensate finite in size and introduce an inhomogeneity across the system, both of which have major implications for the static properties and dynamics of the gases, not least for the bright solitary waves considered herein.  Furthermore, these configurable traps allow for the dimensionality of the system to be engineered to produce ``quasi-one-dimensional" and ``quasi-two-dimensional" systems.

\subsection{Solitons and bright solitary matter waves}

Solitons are non-dispersive waves that arise across nonlinear systems, such as shallow water, plasmas and optical fibres  \cite{ablowitz_segur_book,drazin_johnson_book}.  Although solitons are defined formally as mathematical solutions of nonlinear wave equations,  a physical, ``working" definition is that a soliton \cite{drazin_johnson_book}:
\begin{itemize}
\item Retains its initial shape for all time
\item Is localized
\item Can pass through other solitons and retain its size and shape.
\end{itemize}

The mean-field wave equation of a BEC is of the form of the (3+1)D cubic nonlinear Schr\"odinger equation, with an additional inhomogeneous term arising from the trapping potential.  In the theoretical limit of 1D and in the absence of trapping in the remaining direction, this reduces to the 1D nonlinear Schr\"odinger equation, for which bright and dark soliton solutions are known to exist.  Bright solitons are localized humps in the field amplitude, bound together by a focussing nonlinearity.  Dark solitons appear as localized reductions in an otherwise uniform field amplitude, preserved instead by a defocussing nonlinearity.  While analogs of dark solitons have been observed in BECs (see \cite{frantzeskakis_2010} for a review), we here focus on the case of bright solitons.

Bright soliton-like \cite{strecker_etal_nature_2002,khaykovich_etal_science_2002,cornish_etal_prl_2006} structures have been observed in BECs, with the required focussing nonlinearity arising from the attractive atomic interactions.  Of course, the physical reality introduces three dimensions and trapping potentials/finite-sized systems, and so these are strictly distinct objects from the true bright solitons.  As such we will henceforth refer to this wider family of structures as bright solitary waves.  Following the definition of Morgan {\it et al.} \cite{ morgan_etal_pra_1997} we regard a solitary wave as a wavepacket that propagates without change of shape.  This relaxed definition will allow for the inclusion of solutions which feature trapping potentials and three dimensions, as we shall see.  Of course, one should not assume that a solitary wave will mimic the wider properties of the classic bright soliton and it is a key focus of this Chapter to elucidiate this intimate relationship.

Bose-Einstein condensates are an attractive system in which to study solitonic waves, with some key features summarized below:
\begin{itemize}
\item A sophisticated toolbox based on atomic physics allows almost arbitrary shapes of confining potentials to be constructed, for example, waveguides to steer the wavepackets, systems of reduced dimensionality, and disordered and periodic potential landscapes.
\item  This toolbox also enables the interactions (i.e. the nonlinearity) to be changed effectively from infinitely attractive, through zero, to infinitely repulsive via the exploitation of Feshbach resonances.  Moreover, one can employ atoms such as $^{52}{\rm Cr}$ which feature permanent magnetic dipole moments; this introduces long-range atom-atom interactions, i.e. nonlocal nonlinearity, into the system \cite{lahaye_2009}.
\item The condensate density can be imaged directly with high contrast.  While this is most commonly performed via destructive techniques based on optical absorption, non-destructive imaging techniques are also possible, e.g. phase-contrast imaging \cite{pethick_smith_2002}.  The phase of the condensate can also be mapped out in space and time via interferometric techniques \cite{simsarian_2000}.
\item Bright solitary waves, which typically exist as small BECs, are mesoscopic quantum systems.  This scale allows interfacing of the robust and well-established mean-field description of BECs with more sophisticated models that incorporate thermal and quantum effects \cite{emergent_book}.
\item The precision and control offered by BECs makes them an attractive system in general for application in ultra-precise force detection and quantum information.   For these applications, bright solitary waves offer further merits through their self-trapped, highly-localized form.
\end{itemize}

\subsection{The mean-field Gross-Pitaevskii equation}

Our theoretical analysis will be based upon the well-established Gross-Pitaevskii equation, which is a wave equation for the classical field of the many-body wavefunction \cite{pethick_smith_2002,pitaevskii_stringari_2003,emergent_book}.  This equation is a valid description for a gaseous BEC providing:
\begin{itemize}
\item The condensate is macroscopically-populated, i.e. $N \gg 1$, where $N$ is the number of atoms in the condensate.
\item The temperature of the gas satisfies $T \ll T_{\rm c}$, where $T_{\rm c}$ is the critical temperature for Bose-Einstein condensation, such that approximately all the particles are within the BEC phase.
\item The dominant inter-atomic interactions are two-body, short-range elastic {\it s}-wave collisions, whose lengthscale is parameterized by the {\it s}-wave scattering length $a_{\rm s}$.
\item The condition of length scales  $a_{\rm s} \ll d$, where $d$ the average interparticle distance, is satisfied, such that the detailed shape of the inter-atomic potential becomes unimportant and can be modelled by a simple contact potential (hard-sphere collisions).
\item The interactions are weak, parameterized by the condition $n |a_{\rm s}|^2 \ll 1$, such that fluctuations out of the single-particle state are negligible.
\end{itemize}

Subject to these criteria, the condensate can be parameterized in time and space by a mean-field order parameter $\psi(\mathbf{r,t})$, often termed the {\em macroscopic wavefunction}.    For convenience we take $\psi(\mathbf{r},t)$ to be normalized to unity, i.e.,
\begin{equation}
\int |\psi(\mathbf{r},t)|^2~{\rm d}^3\mathbf{r}=1.
\end{equation}
According to the Madelung transform, $\psi({\bf r},t)$, which is complex, can be related to the atom number density $n({\bf r},t)$ and a phase function $\theta({\bf r},t)$ via,
\begin{equation}
\psi({\bf r},t)=\sqrt{\frac{n({\bf r},t)}{N}}\exp\left[i\theta ({\bf r},t) \right],
\end{equation}
where $N$ is the number of atoms in the condensate (introduced here to account for the normalization of $\psi$ to unity).

The field $\psi$ evolves in space and time according to the Gross-Pitaevskii equation \cite{pethick_smith_2002,pitaevskii_stringari_2003,emergent_book},
\begin{equation}
i\hbar \frac{\partial \psi(\mathbf{r},t)}{\partial t} =
\left[ -\frac{\hbar^2}{2m}\nabla^2
+ V(\mathbf{r})
+\frac{4\pi\hbar^2 N a_s}{m} |\psi(\mathbf{r},t)|^2 \right]
\psi(\mathbf{r},t)\,,
\label{e_3dgpe}
\end{equation}
where $V({\bf r})$ specifies the external potential acting on the condensate (taken, for simplicity, to be time-independent) and $m$ is the atomic mass.

The time-independent eigenstate solutions of Eq.\ (\ref{e_3dgpe}) obey the
GPE in its stationary form,
\begin{equation}
\left[ -\frac{\hbar^2}{2m}\nabla^2
+ V(\mathbf{r})
+ \frac{4\pi\hbar^2 N a_s}{m} |\psi(\mathbf{r})|^2
- \mu \right]
\psi(\mathbf{r})
= 0 \,,
\label{e_3dgpe_static}
\end{equation}
where $\mu$ is a (real) eigenvalue. The lowest energy solution to this
equation represents the mean-field ground state of the BEC.

The Gross-Pitaevskii equation has proven an excellent description of a vast spectrum of static and dynamical properties of BECs \cite{pethick_smith_2002,pitaevskii_stringari_2003,emergent_book}.   The present work will be based primarily on this mean-field description, although we will briefly discuss beyond-mean-field descriptions in Section \ref{s_manybody}.

As is most commonly used to confine a BEC, we will assume a trapping potential that is harmonic in shape.  For simplicity, we will further assume the trap to be cylindrically symmetric.  This restriction sacrifices only a little generality for significant
gains in clarity. We write this potential as,
\begin{equation}
V({\bf r})=\frac{1}{2}m \left[\omega_x^2 x^2 + \omega_r^2(y^2+z^2)\right],
\label{e_trap}
\end{equation}
where $\omega_x$ is the trap frequency in the axial (long) direction and $\omega_r$ is the trap frequency in the transverse directions.  When dealing with such three-dimensional (3D) systems we introduce the trap anisotropy $\lambda=\omega_x/\omega_r$, with $\lambda$ $<$ 1 ($>$ 1) corresponding to
a prolate (oblate) trap.

It is useful to parameterise the interaction strength of the condensate via,
\begin{equation}
k \equiv \frac{|a_s|N}{a_r}\,,
\label{e_kdefgeneral}
\end{equation}
where $a_r = \sqrt{\hbar/m\omega_r}$ is the harmonic oscillator length in the
radial direction.\footnote{Note that in works that focus specifically on fully trapped condensates, $k$
is more commonly defined in terms of a geometric average of trap frequencies (e.g.
Refs.\ \cite{roberts_etal_prl_2001, cornish_etal_prl_2006,
gammal_etal_pra_2001, gammal_etal_pra_2002, huepe_etal_prl_1999,
berge_etal_pra_2000, akhmediev_etal_ijmpb_1999, houbiers_stoof_pra_1996,
adhikari_njp_2003, dodd_etal_pra_1996, yukalov_yukalova_pra_2005}). The radial harmonic oscillator length here
is advantageous as it allows us to readily consider the case of zero axial trapping
($\lambda=0$).}

\subsection{Chapter overview}

The bright solitary waves generated experimentally are related, but strictly different, entities to the true bright solitons (which apply only in 1D and for a uniform, infinite system).  It is the focus of this Chapter to explore this relationship in detail, highlighting the similarities and differences.  In essence, we wish to shed light on how ``soliton-like" these solitary waves are.  We will consider how the waves look, how they move and how they interact with each other.  The deviation of the bright solitary wave from the true bright soliton is a consequence of two factors: the inclusion of an inhomogeneous trapping potential and the extension to three-dimensions, and we will consider these two factors separately so as to elucidate their independent contributions to the identity of bright solitary waves.  We first begin in Section \ref{s_experiment} by outlining the experimental generation and observation of bright solitary matter waves to date.   Following this we begin our theoretical analysis of bright solitary waves.  Sections \ref{s_static1d} and \ref{s_static3d} explore the static solutions of bright solitary waves.  In Section \ref{s_static1d} this is conducted within an effective 1D model of the condensate, and the role of axial trapping considered.  Then in Section \ref{s_static3d} we extend our analysis of the static solutions to 3D, where the collapse instability comes into play.  In Sections  \ref{s_dynamic1d} and \ref{s_dynamic3d}  we turn to the dynamics of bright solitary waves in 1D and 3D, respectively.  There we consider the solitary wave dynamics resulting from the presence of axial trapping and the interaction with another solitary wave.  In Section \ref{s_outlook} we turn our attention to the state-of-the-art in bright solitary wave research, discussing beyond-mean-field descriptions and the current anomalies with mean-field predictions, observations and predictions of more exotic bright solitary waves, and proposals for controllable generation of bright solitary waves and exploiting them as ultra-precise atom-optical sensors.  Finally, in Section \ref{s_conclusions}, we draw some general conclusions.

\section{Bright solitary matter wave experiments}
\label{s_experiment}

In order to experimentally realise bright solitary matter waves precise control over the $s$-wave scattering properties of an atomic sample is of paramount importance. In the following section we discuss the application of magnetic Feshbach resonances as a means of establishing this control and review bright solitary matter wave experiments to date.

\subsection{Tuning atomic interactions: Feshbach resonances}

\begin{figure}
	\centering
		\includegraphics[width=\textwidth]{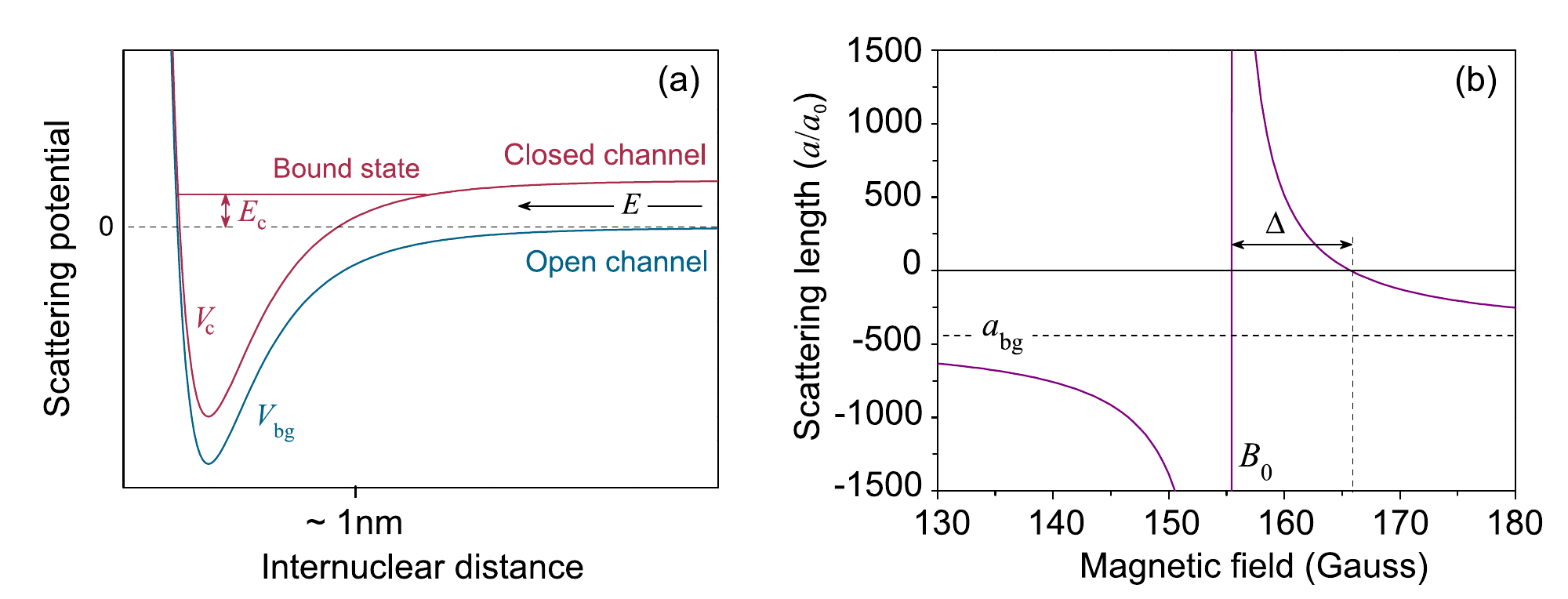}
	\caption{\small Feshbach resonances: (a) A two channel model of a Feshbach resonance. A resonance occurs when two atoms colliding with energy $E$ resonantly couple to a bound state of the closed channel. (b) The Feshbach resonance present in the $F=2, m_F=-2$ state of $^{85}$Rb. }
	\label{fig:Feshbach_resonance_and_85}
\end{figure}

The use of magnetically tunable Feshbach resonances \cite{chin_etal_rmp_2010} to control the interaction between atoms is now commonplace in many ultracold atomic gas experiments. Feshbach resonances arise when a resonant coupling occurs between the collisional open and closed channels of an atomic system, as illustrated in Fig. \ref{fig:Feshbach_resonance_and_85}(a).  For large internuclear distances, the interaction between two atoms can be described by the background potential, $V_{\rm{bg}}$. If two free atoms approach, colliding with low energy, $E$, this potential represents the open or entrance channel for the collision. In contrast, closed channels (described by $V_{\rm{c}}$) are able to support molecular bound states. A Feshbach resonances occurs when the energy of a quasi-bound molecular state in the closed channel, $E_{\rm{c}}$, approaches that of the open channel. In this instance a strong mixing between the two channels can occur even in the presence of only weak coupling. By changing the magnetic field applied this energy difference can be tuned if the magnetic moments of the two channels differ thus the scattering properties of the atomic sample can be modified.

These resonances allow the value of the $s$-wave scattering length, $a_s$, to be changed over many orders of magnitude in both the positive and negative domain by simply changing the magnetic field,
\begin{equation}
a_s(B)=a_{\rm{bg}}\left(1-\frac{\Delta}{B-B_0}\right).
\label{eq:}
\end{equation}
Here $a_s$ is the scattering length at the field of interest, $B$, $a_{\rm{bg}}$ is the background scattering length away from the resonance, $\Delta$ is the width of the resonance and $B_0$ is the resonance position. In the case
of broad resonances, where $\Delta \gtrsim$ 1 G, there is a smooth variation of the scattering length
through zero from positive to negative with a slope of $da/dB = a_s/\Delta$. For Bose-Einstein condensation of some species (e.g. $^{85}$Rb, $^7$Li) this is of particular importance as it allows the creation of stable condensates with repulsive interactions ($a_s>0$) despite a negative background scattering length away from the resonance. As an illustration, Fig. \ref{fig:Feshbach_resonance_and_85}(b) shows the Feshbach resonance in the $F=2, m_F=-2$ state of $^{85}$Rb. This broad resonance, of width $10.7$~G, at $\sim$155~G gives tuning of the scattering length on the order $40a_0$/G close to the zero crossing.

In all of the experiments described in the following sections a Feshbach resonance is the key atomic tool without which the controlled creation of bright solitary matter waves would not be possible.


\subsection{Collapse of an attractive Bose-Einstein condensate}

A BEC (in three-dimensions) with attractive interactions is inherently unstable to collapse when its interaction parameter $k=N |a_s|/a_r$ exceeds a critical value $k_{\rm c}$.  This leads to the typical notion of a critical atom number $N_{\rm{c}}$ (for fixed $a_s$ and $a_r$) or critical scattering length $a_c$ (for fixed $N$ and $a_r$) at which instability becomes induced.  The origin of the collapse instability will be outlined theoretically in Section \ref{s_collapse_parameter}. The ensuing collapse of the condensate has been dubbed the `Bosenova' in analogy to the astronomical phenomena of stellar explosion.

The first experimental insights into BECs with attractive interactions were made using ${^7}$Li \cite{bradley_etal_prl_1997}. Here the negative scattering length of $a_s=(-27.4\pm0.8)~a_0$, where $a_0=5.29 \times 10^{-11}~{\rm m}$ is the Bohr radius, means that the condensate atom number $N$ grows until it reaches $N_c$ and the condensate collapses. During the collapse the density of the cloud rises thus increasing both the elastic and inelastic collision rates. This causes atoms to be ejected from the condensate with high energy in a violent explosion. Following this, the condensate begins to reform, fed by the surrounding bath of thermal atoms also present in the trap. If observed for an extended period the system exhibits a saw-tooth dynamic of growth and collapse \cite{gerton_etal_nature_2000} until equilibrium is eventually reached. Throughout, the maximum condensate number is strictly limited to the critical number for an attractive BEC. It is also possible that collapse occurs even with $N<N_{\rm{c}}$ due to quantum tunnelling effects and thermal fluctuations in the cloud leading to instability.

Further insight into the collapse phenomena came from the group at JILA (Boulder, US) in 2001 \cite{donley_etal_nature_2001, roberts_etal_prl_2001}, carrying out a controlled collapse using a pure $^{85}$Rb condensate. Tuning the scattering length from positive to negative using the broad Feshbach resonance illustrated in Figure \ref{fig:Feshbach_resonance_and_85} not only enabled the collapse process to be precisely initiated but also allowed the condition $k> k_{\rm{c}}$ to be fulfilled, unlike systems using fixed negative scattering lengths. Along with control of the initial condensate number, control over scattering length made the testing of critical number models possible, finding the exact scattering length necessary to collapse the cloud, $a_{\rm{c}}$. Early work examining the point of collapse using slow field ramps confirmed the relationship between critical number and scattering length, determining the critical interaction parameter for $k_c$. Later improvements to the calibration of the Feshbach resonance, enhancing precision, found $k_c$ to be in excellent agreement with mean-field models \cite{PhysRevA.67.060701}.

\begin{figure}
	\centering
		\includegraphics[width=0.6\textwidth]{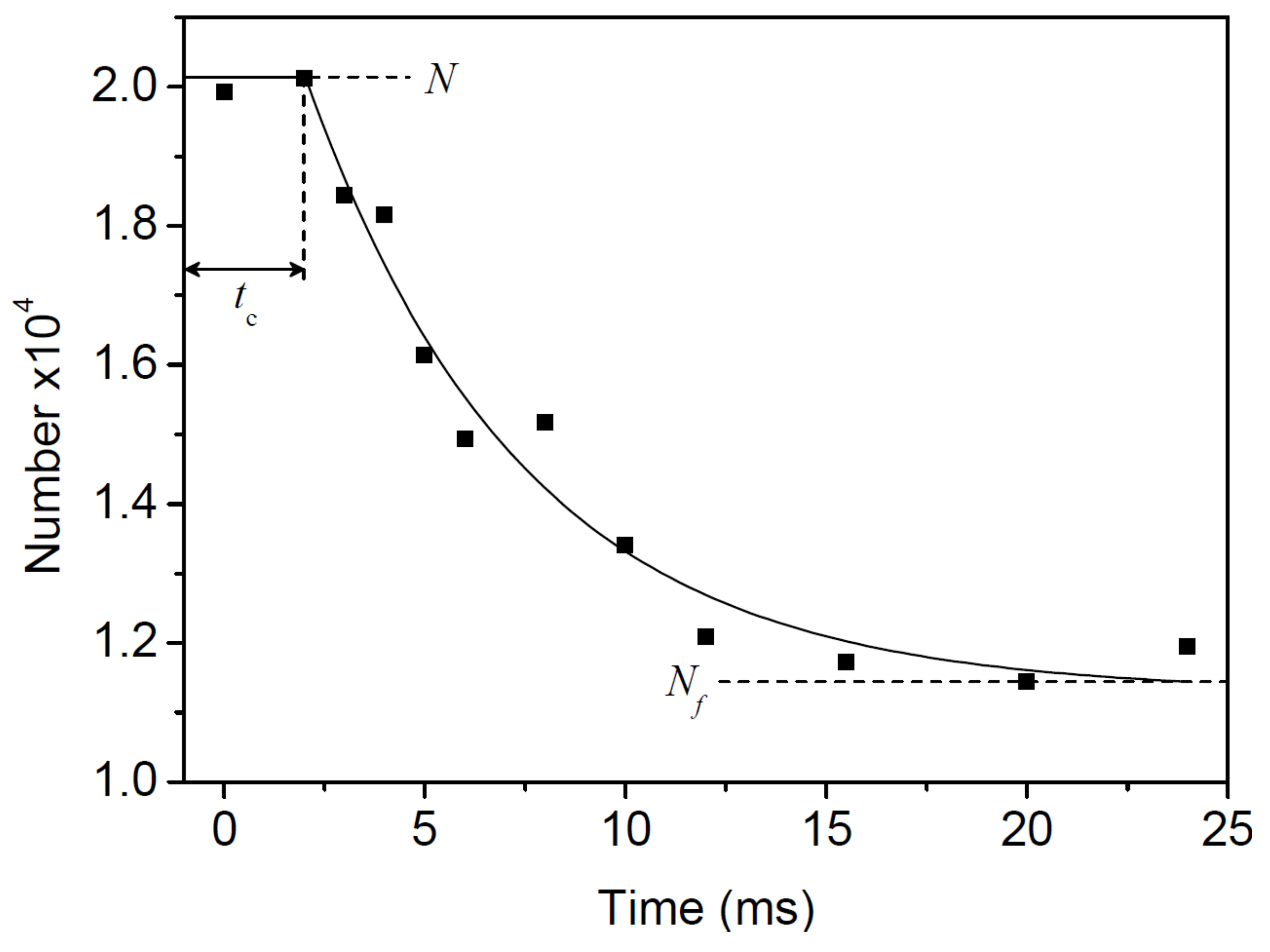}
	\caption{\small Controlled collapse: The collapse of a stable Bose-Einstein condensate can be triggered by a sudden change of the scattering length from positive to negative. After some time at the new scattering length, $t_{\rm{collapse}}$, the condensate begins to collapse and atoms are lost. Eventually the collapse process ceases, leaving a stable remnant in the trap containing $N_f$ atoms. [Data from Durham $^{85}$Rb experiment] }
	\label{fig:Collapse}
\end{figure}

Following this, the JILA experiments were then extended to study the dynamics of the collapse, measuring the evolution of the condensate number following a `sudden' change in the scattering length. Measurements of atom number as a function of time showed a sudden yet delayed loss of atoms, as shown in Fig. \ref{fig:Collapse}. As the interactions are made attractive the condensate begins to shrink in size, thus increasing its density. This contraction tends to accelerate with time eventually leading to collapse of the condensate. The time for the collapse to begin $t_{\rm{c}}$ was found to be shorter for larger $\left|a_{\rm{s}}\right|$ as the stronger attraction between atoms in the condensate results in a more rapid contraction of the cloud. Following the collapse, a stable remnant component was formed in the trap. Notably, the number of atoms maintained in this remnant, $N_f$, was found to depend strongly on $N$ and $a_{\rm{s}}$ and in many cases exceeded $N_{\rm{c}}$. This remnant was observed to persist in the trap for more than 1~s, oscillating in a highly excited state.

In addition, a number of more qualitative features were observed about the collapse process in the $^{85}$Rb experiment. The first of these features was bursts of atoms with variable energies being ejected from the condensate. These bursts would then focus at multiples of $T_{x}/2$ and $T_{r}/2$, where $T_{x,r}=2\pi/\omega_{x,r}$ is the trap period in the axial ($x$) and radial ($r$) dimensions. In all experiments only full, never partial, collapse was observed. However, if interrupted (by jumping the scattering length away from the collapse point), jets of atoms were also formed. Unlike the bursts, these streams of atoms were found to have highly anisotropic velocities and were interpreted as indicating the local pinching of the wavefunction during the collapse.

The collapse process has since been revisited by the group at the Australian National University (Canberra, Australia) \cite{ altin_etal_pra_2011}. Again using $^{85}$Rb, measurements of the collapse time have been shown to be in good agreement with mean-field models describing the process which take into account three-body loss mechanisms.


\subsection{Observation of bright solitary matter waves}

The advent of optical trapping led to the realisation of experimental geometries closer to the ideal 1D limit. This, in combination with control of the atomic scattering length via Feshbach resonances, led to the first observations of bright solitary matter waves by groups at Rice University (Houston, US) \cite{strecker_etal_nature_2002} and Ecole Normale Suprieure (Paris, France) \cite{khaykovich_etal_science_2002} in 2002 using $^{7}$Li. Despite two inherently similar experiments, the ENS group succeeded in producing a single solitary wave whereas the Rice experiment resulted in trains of multiple solitary waves.

In order to utilize the Feshbach resonance in the non-magnetically trappable $F=1, m_F=1$ state of $^{7}$Li it is necessary to work using an optical dipole trap \cite{Grimm200095}. In both experiments initial cooling of the atomic sample was carried out in a magnetic trap using the $F=2, m_F=2$ state before transferring to a dipole trap and flipping the spin state of the atoms to suppress two-body loss mechanisms and allow access to the Feshbach resonance.

In the ENS experiment optical confinement was realised using a red-detuned crossed dipole trap. Here condensates of $2\times10^4$ atoms were produced with $a_s=+39.7a_0$ in an approximately isotropic harmonic trap. After the creation of the BEC the scattering length was tuned close to $a_s=0$ before adiabatically reducing the power in one of the beams, producing a highly elongated cylindrical harmonic trap with $\omega_x=2\pi\times 50$~Hz and  $\omega_r=2\pi\times 710$~Hz. The bias field, and hence scattering length, was then ramped to its final value before the vertical beam was switched off, releasing the cloud into a 1D waveguide. In this trap, the atoms experience a slightly expulsive potential due to the magnetic coils used to produce the bias field. As a typical example, at $B=520$~G, the trap frequency along the waveguide can be considered imaginary, around $\omega_x=2{\rm i}\pi\times78$~Hz. Tuning the scattering length to a small negative value, $a_s=-3.97~a_0$, resulted in a soliton of $6\times10^3$ atoms able to propagate without dispersion for over 1.1~mm.

In contrast to the crossed ENS trap, the Rice experiment used a single red-detuned dipole beam to provide radial confinement.  Two additional blue detuned beams were applied to cap the ends of the trap in the axial direction. After forming a condensate of $3\times10^5$ atoms with $a_s\approx200a_0$ the magnetic field controlling the scattering length was ramped exponentially to the final value and the laser end caps switched off thus setting the resulting solitary waves in motion.

In this experiment multiple solitary waves were observed. The number of these wavepackets, $N_s$, was found to be insensitive to the time constant of the exponential magnetic field ramp.  However, $N_s$ increased linearly with $\Delta t$, the time delay between the switch off of the end caps and the time of the scattering length change to $a_s<0$. For the Rice experiment four solitary waves were observed for $\Delta t=0$ with this number increasing to 10 for $\Delta t=35$~ms. The wavepackets were observed to propagate for $\sim$3~s, this being limited by atom loss rather than dispersion effects.

With many solitary waves confined in a single trap it becomes possible to explore the dynamical interactions of the wavepackets. Observation of the solitary wave motion showed evidence of a short range repulsive interaction between neighbouring wavepackets raising many questions regarding their formation and collisional dynamics. A possible explanation for the formation of multiple solitary waves was the presence of a modulational instability \cite{PhysRevLett.56.135}. Here, phase fluctuations of the condensate lead to a local increase in density at wavelengths approximately equal to the healing length. The attractive nonlinearity leads to the growth of these density fluctuations and the emergence of solitons.

The spacing between neighbouring solitary waves observed at Rice increased near the centre of the oscillation and decreased near the turning points. This result implied a repulsive interaction between solitary waves.  This interaction was attributed to the existence of $\pi$-phase differences between neighbouring solitary waves, somehow imprinted during their formation.  The phase-dependence of the solitary wave interaction will be discussed in Sections \ref{s_nlse_dynamics} and \ref{s_dynamic3d}, and the origin of the $\pi$-phase difference in Section \ref{s_manybody} .

It was not until 2006 that solitary waves were again investigated experimentally, this time at JILA  \cite{cornish_etal_prl_2006} using the same $^{85}$Rb experiment that has first observed tunable atomic interactions \cite{PhysRevLett.85.1795} and controlled collapse \cite{roberts_etal_prl_2001}. This new work concluded that the stable remnant observed previously in the collapse experiments divided into similar solitary wave structures as seen at Rice.  Intriguingly, these observations persisted despite the fact that the JILA trap remained almost isotropic (with radial and axial trap frequencies of 17.3~Hz and 6.8~Hz respectively), far from the highly elongated geometries employed at ENS and Rice.  The somewhat surprising capacity of bright solitary waves to be supported in almost isotropic trap geometries will be discussed in Section \ref{s_ground_state_properties}.

Unlike the ENS and Rice experiments, the JILA apparatus used a purely magnetic trap.  However, the method of creating solitary waves by modifying the scattering length can be considered an inherently similar process. After producing condensates of up to 15,000 atoms at $a_s>0$ the magnetic field was adiabatically ramped to decrease the scattering length to $a_s=9a_0$. To initiate the collapse, the magnetic field was changed so as to rapidly ($0.1$ ms) jump the scattering length from a positive initial value to a negative final value of $a_{\rm f}$. Following some time at the final scattering length, $t_{\rm{evolve}}$, the atoms were destructively imaged. Investigating the collapse process as a function of $a_{\rm{f}}$ and the initial condensate number it was clear that the number of condensate atoms surviving the collapse could greatly exceed $N_{\rm{c}}$.  Furthermore, the lifetime of the stable remnant could be as long as several seconds.  As we will review in Section \ref{s_dynamic3d_trap}, this observation is consistent with the presence of several repulsively-interacting bright solitary waves.

\begin{figure}[t]
	\centering
		\includegraphics[width=0.7\textwidth]{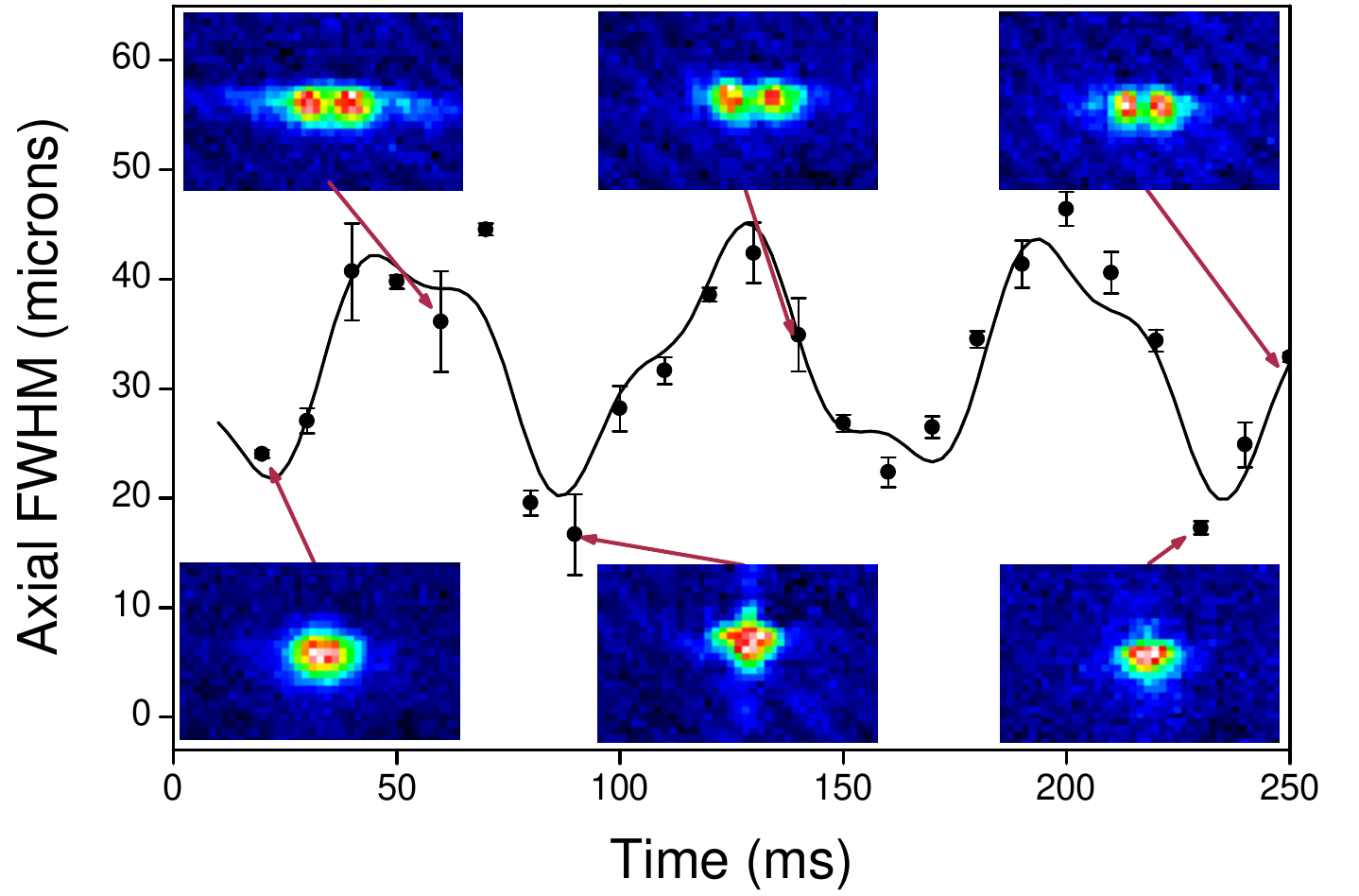}
	\caption{\small Solitary wave oscillation in a weak magnetic trap \cite{cornish_etal_prl_2006}: Following the collapse process used in the JILA experiment a stable remnant is formed. The variation in the remnant's width with time can be explained by the creation of multiple bright solitary matter waves oscillating in the trap, which are visible in the 2D plots of atomic density (insets).}
	\label{fig:SolitonOscillation}
\end{figure}

Observations of the condensate size in the trap as a function of time suggested a highly excited state had been produced during the collapse, with the remnant cloud's width doubling during its oscillation in the trap. However, further analysis revealed that, as in the Rice experiment, multiple solitary waves were being created which oscillated back and forth along the weak axial direction of the trap, shown in Fig. \ref{fig:SolitonOscillation}. The wavepackets were observed to persist in the trap for $\sim3$~s, undergoing as many as 40 collisions in this time. This provided additional experimental data to accompany the Rice experiments and the growing body of theoretical work on the stability of three-dimensional bright solitary waves (which we will review in Sections \ref{s_static3d} and \ref{s_dynamic3d}). The number of solitary waves created in the $^{85}$Rb collapse experiment was found to be controllable, to a degree, depending on $a_{\rm{f}}$ and $N_0$. As expected, $N_s$ increased with $|a_{\rm{f}}|$. Importantly, the number of atoms observed in any one solitary wave was never found to exceed $N_{\rm{c}}$.

\subsection{Current developments}

In order to further explore the results from both previous experiments and theoretical simulations, an experiment has been constructed at Durham University (Durham, UK). As in the JILA experiment, this apparatus uses $^{85}$Rb in the $F=2, m_F=-2$ state allowing access to the 10.7~G wide Feshbach resonance giving control over the scattering length of order $40~a_0/$G close to $a=0$. However, the trapping geometry, a crossed dipole trap and additional waveguide beam to produce a quasi 1D geometry, allows entirely independent control of the trap frequencies and $s$-wave scattering properties (due to an independent magnetic bias field).

\begin{figure}
	\centering
		\includegraphics[width=0.9\textwidth]{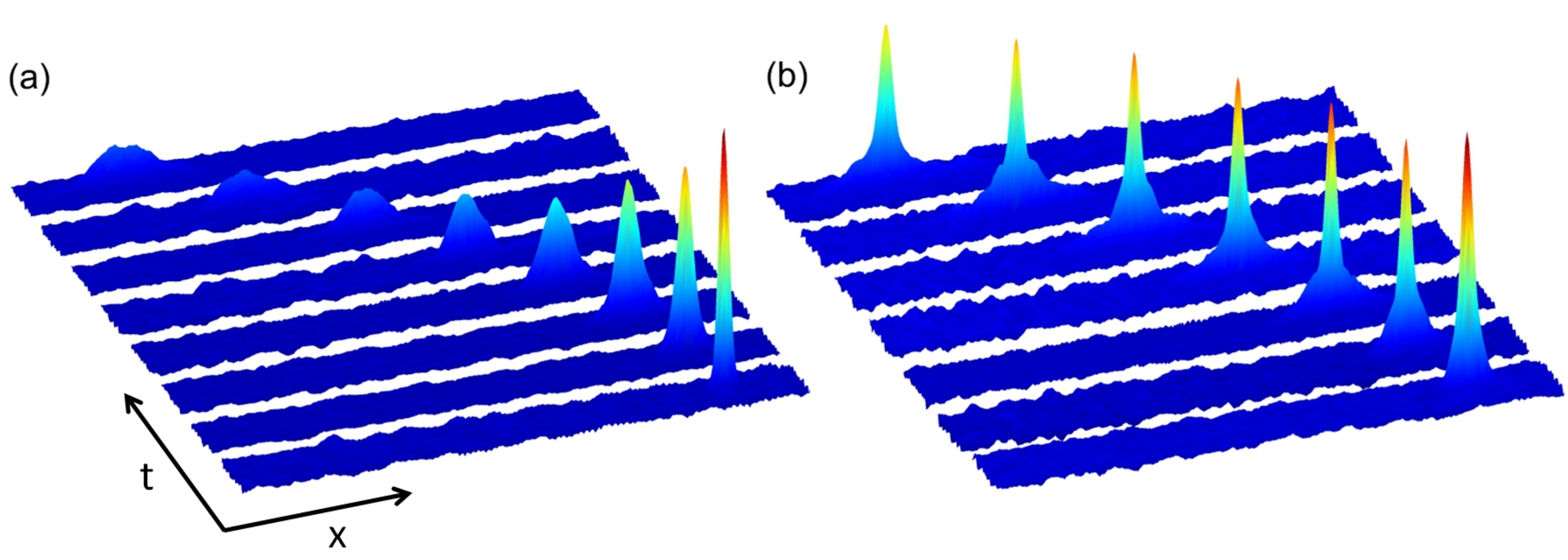}
	\caption{\small Propagation in the waveguide: (a) As a repulsive BEC travels along the waveguide the interactions in the condensate cause it to spread out. (b) In contrast, the attractive interactions present in a bright solitary matter wave cause the wavepacket to hold together as it propagates, maintaining its shape with time. [Data from Durham $^{85}$Rb experiment]  }
	\label{fig:Propagation}
\end{figure}

Here BECs are first created in the crossed dipole trap (at $300-400~a_0$) by careful tuning of the atomic scattering properties. Once condensed, the scattering length is ramped close to $a=0$ before the BEC is loaded into the waveguide by synchronously switching the cross beams off and the waveguide beam on. The scattering length is then jumped to a small, negative value ($\sim-6.5~a_0$) and the BEC is allowed to propagate along the waveguide as shown in Fig. \ref{fig:Propagation}. Weak axial confinement along the beam is realised with the addition of a magnetic quadrupole gradient. Typically this results in trap frequencies of $\omega_x=2\pi\times 1$~Hz and $\omega_r=2\pi\times 27$~Hz. Using this method a single soliton can be produced containing $\sim$3,000 atoms observed to propagate a distance of $\sim$1~mm in 150~ms.

In addition to experiments aimed at investigating soliton splitting and binary collisions (the theory of which will be detailed in Section \ref{s_split_interf}), the Durham experiment has the potential to be extended to the study of atom-surface interactions. Contained within the experimental apparatus is a super polished Dove prism (surface roughness $<1~$\AA) designed to allow the study of both classical and quantum reflection from a surface. The self-stabilizing, localized nature of the wave packets means bright solitary matter waves show great potential as surface probes for the study of short-range atom-surface interactions in the future.   This idea will be explored in more detail in Section \ref{s_surface}.

The Rice group have reported further experimental activity on bright solitary matter waves \cite{dyke_2011}.  Here they ``'kick" a bright solitary wave towards a thin potential barrier, formed by a near-resonant focussed laser beam.   The wave-barrier interaction is observed to result in either reflection, transmission and splitting of the solitary wave, depending on the kinetic energy of the solitary wave, the potential strength and the nonlinearity ({\it s}-wave interaction strength).   Moreover, for the case of a split solitary wave, they have applied a phase imprinting to one of the solitary waves and thereby studied phase-dependent interaction of solitary waves \cite{dyke_2012}.


\section{Bright solitary waves in 1D: static properties}
\label{s_static1d}

Having reviewed the experimental formation and observation of bright solitary matter waves to date, we will now review our theoretical understanding of these wavepackets.  Within a quasi-one-dimensional (quasi-1D) system, bright solitary
matter waves become completely analogous --- within the mean-field,
Gross-Pitaevskii equation (GPE) treatment --- to classical bright solitons of the 1D nonlinear
Schr\"{o}dinger equation (NLSE) \cite{dauxois_book}. In this section we examine
the quasi-1D limit in which this occurs. In Section \ref{s_quasi1dgpe} we
describe the conventional factorization to reduce the 3D GPE to an effective 1D form, and some
approaches to include higher-order terms. In Section
\ref{s_brightsolitons} we demonstrate the link to bright
solitons and explore the static properties of bright soliton solutions of the
NLSE. Then, in Section \ref{s_axialtrapping}, we consider the bright solitary
matter waves which occur as the ground state of an axially trapped
BEC, elucidating how their form depends on the strength of the axial
trap, and how they compare to the bright solitons of the 1D NLSE.

\subsection{Effective one-dimensional descriptions}
\label{s_quasi1dgpe}

We begin by considering an attractively-interacting ($s$-wave scattering
length $a_s < 0$) three-dimensional BEC confined by the cylindrically-symmetric harmonic trap of Eq. (\ref{e_trap}) and  described by the 3D Gross-Pitaevskii equation, Eq. (\ref{e_3dgpe}).

\subsubsection{Quasi-one-dimensional GPE}
\label{s_std_reduction}

The quasi-1D limit is associated with highly elongated ($\omega_r \gg \omega_x$)
traps.  The reduction from the full 3D to an effective 1D description typically proceeds by assuming that the radial confinement is sufficiently strong.\footnote{Specifically, the criteria $\hbar \omega_r \gg \mu$ and $\hbar \omega_r \gg k_{\rm B}T$
are required to ensure that the condensate and thermal energy scales are insufficient to excite the radial modes.} that the
radial modes of the condensate become essentially ``frozen'' into the ground
harmonic oscillator ground state (i.e. a Gaussian wavefunction)  This approximation then allows the factorization,
\begin{equation}
\psi(\mathbf{r}) = \sqrt{\frac{m \omega_r}{\pi \hbar}}
\exp\left[\frac{-m\omega_r (y^2 + z^2)}{2 \hbar}\right]
\psi(x)\,
\label{e_factorization}
\end{equation}
where it is implied that both the Gaussian radial wavefunction and the axial wavefunction are both normalized to unity.
Integrating over the $y$- and $z$-directions then yields the quasi-1D GPE for
$\psi(x)$,
\begin{equation}
i \hbar \frac{\partial \psi(x,t)}{\partial t} =
\left[ -\frac{\hbar^2}{2m}\frac{\partial^2}{\partial x^2}
+ \frac{m\omega_x^2 x^2}{2}
- 2 \hbar \omega_r  N |a_s| |\psi(x,t)|^2 \right]
\psi(x,t)\,.
\label{e_q1dgpe}
\end{equation}
In the static case, one obtains the stationary quasi-1D GPE,
\begin{equation}
\left[ -\frac{\hbar^2}{2m}\frac{\partial^2}{\partial x^2}
+ \frac{m\omega_x^2 x^2}{2}
- 2 \hbar \omega_r  N |a_s| |\psi(x)|^2
-\mu \right]
\psi(x)
=0\,.
\label{e_q1dgpe_static}
\end{equation}
This factorization has often been applied in the study
of attractively-interacting condensates (in both dynamic and static situations)
\cite{perez-garcia_etal_pra_1998, al_khawaja_etal_prl_2002,
carr_brand_prl_2004, leung_etal_pra_2002, martin_etal_prl_2007,
martin_etal_pra_2008}. However, the regime in which this factorization is valid
is significantly restricted for attractively-interacting condensates
\cite{billam_etal_variational_2011}; this issue is revisited using a full 3D
analysis in Section \ref{s_static3d}.

\subsubsection{One-dimensional equations with 3D effects}
\label{s_extra_reduction}

Alternatives to the factorization presented above exist, which yield 1D
equations retaining more 3D character by choosing to incorporate the coupling
between axial and radial modes, and time-dependent dynamics of the radial modes
\cite{salasnich_etal_pra_2002a, salasnich_etal_pra_2002b,
kamchatnov_shchesnovich_pra_2004, khaykovich_malomed_pra_2006}. These effects
are manifest in the effective 1D equation through the appearance of
higher-order terms. Consequently, the resulting equations have a wider range of
validity than the bare 1D GPE (\ref{e_q1dgpe}), but are no longer
isomorphous to the NLSE (for $\omega_x =0$).

For example, Salasnich \textit{et al.} \cite{salasnich_etal_pra_2002a,
salasnich_etal_pra_2002b} chose to factorize the 3D GPE wavefunction into a
slowly-varying axial function, multiplied by a rapidly varying radial function.
The radial function was also given a dependence on the axial function itself;
this incorporates the effect unique to attractive interactions in a
cigar-shaped trap, where an increase in axial density leads to an associated
increase in radial density. A variational calculation then yields
the non-polynomial Schr\"{o}dinger equation \cite{salasnich_etal_pra_2002a},
\begin{eqnarray}
\label{e_npse}
i \hbar \frac{\partial \psi(x,t)}{\partial t} &=&
-\frac{\hbar^2}{2m} \frac{\partial^2 \psi(x,t)}{\partial x^2}
+ \frac{m \omega_x^2 x^2}{2} \psi(x,t)
+ \frac{2 \hbar^2 |a_s| N |\psi(x,t)|^2\psi(x,t)}{m a_r \sqrt{1-2|a_s|N|\psi(x,t)|^2}} \\
& & \mbox{} + \frac{\hbar \omega_x}{2}\left(
\frac{1}{\sqrt{1-2|a_s|N|\psi(x)|^2}}+\sqrt{1-2|a_s|N|\psi(x,t)|^2}
\right) \psi(x,t)\,.\nonumber
\end{eqnarray}
When $|a_s|N|\psi(x)|^2 \ll 1$ for all $x$ this reduces first to an effective
1D equation with both cubic and quintic nonlinearities
\cite{khaykovich_malomed_pra_2006}, and then to the bare 1D GPE  (\ref{e_q1dgpe}) itself. An
even more general approach can be taken, incorporating even fewer assumptions
about the form of the ground state, but leading to a coupled system of
effective 1D equations \cite{kamchatnov_shchesnovich_pra_2004}.

\subsection{Bright soliton solutions}
\label{s_brightsolitons}

Consider the 1D GPE (\ref{e_q1dgpe}) in the homogeneous regime $\omega_x = 0$.  With the removal of the quadratic potential term, this becomes a 1D nonlinear Schr\"{o}dinger equation (NLSE) with a focusing
nonlinearity \cite{dauxois_book}.  The 1D NLSE is a classical field equation
which is integrable, in the sense that solutions possess an infinite and
complete set of conserved quantities \cite{dauxois_book, fadeev,
ablowitz_segur_book}. This is analogous to a discrete system which possesses as
many conserved quantities as it does degrees of freedom \cite{hand_finch}. This
integrability leads to a spectrum of true soliton solutions \cite{fadeev,
ablowitz_segur_book}.  In the case of the 1D NLSE with focusing nonlinearity,
these bright-soliton solutions were first discovered in Ref.\
\cite{zakharov_shabat_1972, zakharov_shabat_1972_russian} using the inverse
scattering technique (see Refs.\ \cite{fadeev, ablowitz_segur_book} for an
overview).

The classical bright soliton solutions of this equation have been extensively
studied in the context of optical solitons \cite{zakharov_shabat_1972,
zakharov_shabat_1972_russian, satsuma_yajima_1974, gordon_ol_1983,
kodama_hasegawa_1991, afanasjev_vysloukh_j_opt_soc_am_b_1994,
haus_wong_rmp_1996}.  The same equation appears in many other fields, including
biophysics, astrophysics and particle physics \cite{dauxois_book}, and in the
study of deep ocean waves \cite{akhmediev_etal_pla_2009}. The single-bright-soliton solution of the homogeneous 1D GPE is given by,
\begin{eqnarray}
\label{e_generalsoliton}
\psi(x,t) &=& \frac{a}{2\sqrt{b_x}}
\mbox{sech}\left[\frac{a(x-x_0-v t)}{2b_x} \right]\\
& & \mbox{} \times \exp\left[
i\left\{
\frac{m}{\hbar} \left(v(x-x_0)
+ \frac{v^2 t}{2}
+ \frac{\omega_r^2 |a_s|^2 N^2 a^2 t}{2} \right)
+ \Phi
\right\} \right]\,.\nonumber
\end{eqnarray}
This solution describes a single bright soliton with amplitude and
norm\footnote{In contrast to our definition here, a common convention in the
literature is to define an amplitude $A$ such that the norm is $2A$
\cite{gordon_ol_1983}} $a$, velocity $v$, displacement $x_0$, and phase $\Phi$.
The parameter $b_{x} \equiv \hbar / 2 m \omega_r |a_s| N$ is a length scale
characterizing the soliton's spatial extent.
Dynamical solutions composed of
multiple bright solitons also exist; in these solutions each soliton has a
similar form to Eq.\ (\ref{e_generalsoliton}) when well-separated from the
others.
These
multiple-soliton solutions contain additional, dynamic phase and position
shifts to account for the nonlinear interactions between solitons; these
dynamical solutions are discussed further in Section \ref{s_dynamic1d}.


The single-soliton ground state of the static 1D GPE (\ref{e_q1dgpe_static}) in its homogeneous ($\omega_x=0$) form is given
exactly by Eq.  (\ref{e_generalsoliton}) with $a=1$, $v=0$, and arbitrary
$\Phi$ and $x_0$. The quantity $\Phi$ can be chosen arbitrarily because it
corresponds to a global phase of the wavefunction, and Eq.\
(\ref{e_q1dgpe_static}) possesses a $U(1)$ global phase symmetry. Similarly, the
displacement $x_0$ may be chosen arbitrarily because the assumption of homogeneity ($\omega_x=0$) ensures the 1D GPE to be translationally symmetric. However, the choice of
displacement $x_0$ in Eq.\ (\ref{e_generalsoliton}) for the ground state breaks
this symmetry; in the context of atomic BECs, this symmetry-breaking is a
feature of the mean-field description. This feature is at odds with a fully
quantum-mechanical treatment; in the latter, the ground state of the system
retains the translational symmetry of the equation, leading to a delocalized
ground state \cite{mcguire_1964,holdaway_pra_2012}.

\subsection{Effect of axial trapping}
\label{s_axialtrapping}
The addition of an axial harmonic trap ($\omega_x>0$) removes the integrability of the
system and prevents the appearance of true solitons. While the new ground state is no longer a soliton, it remains
a solitary wave in the sense of being capable of propagation without dispersion
(see Section \ref{s_dynamic1d} and Refs. \cite{morgan_etal_pra_1997,
martin_etal_prl_2007, martin_etal_pra_2008}).  In this section we elucidate the
form of the ground state under axial trapping and compare the form of this ground state to the NLSE
bright soliton.

\subsubsection{Variational analysis}
A great deal of insight into the form of the ground state can be gained using a
variational approach \cite{billam_etal_variational_2011}; such approaches have
proved useful for treating a variety of problems involving bright solitary
matter waves \cite{malomed_po_2002,  carr_castin_pra_2002,
parker_etal_jpb_2007, perez-garcia_etal_prl_1996, perez-garcia_etal_pra_1997,
salasnich_etal_pra_2002a, salasnich_etal_pra_2002b, ernst_brand_pra_2010,
khaykovich_malomed_pra_2006, kamchatnov_shchesnovich_pra_2004} and will be used
extensively in Section \ref{s_static3d}.

The ground state solution of Eq. (\ref{e_q1dgpe_static}) can be alternatively
defined as the function $\psi(x)$ which minimizes the value of the classical
field Hamiltonian,
\begin{equation}
H_\mathrm{1D}[\psi(x)] = \int dx \left[
\frac{\hbar^2}{2m}\left|\frac{\partial}{\partial x}\psi(x)\right|^2
+ \frac{m\omega_x^2 x^2}{2}|\psi(x)|^2
- \hbar \omega_r  N |a_s| |\psi(x)|^4 \right]\,.
\label{e_1dhamiltonian}
\end{equation}
This functional represents the energy per particle, and generates the 1D
GPE (\ref{e_q1dgpe}) through the functional derivative $\delta
H_\mathrm{1D}[\psi]/\delta \psi^\ast = i \partial \psi / \partial t$.

In the homogeneous limit ($\omega_x=0$) the ground state is given by Eq.
(\ref{e_generalsoliton}) with $a=1$ and $v=0$. In the trap-dominated limit
($\omega_x \rightarrow \infty$) the ground state tends to the harmonic
oscillator eigenstate $\psi(x) = (m\omega_x/\pi\hbar)^{1/4}e^{-m\omega_x
x^2/2\hbar}$. These limits motivate our use of a normalized Gaussian ansatz,
\begin{equation}
\psi(x) =
\left(\frac{m\omega_x}{\pi\hbar\ell_{x,\mathrm{G}}^2} \right)^{1/4}
\exp\left(-\frac{m\omega_x x^2}{2\hbar \ell_{x,\mathrm{G}}^2}\right)\,,
\label{e_1dgaussansatz}
\end{equation}
or a normalized sech ansatz,
\begin{equation}
\psi(x) =
\frac{1}{2\sqrt{b_x \ell_{x,\mathrm{S}}}}
\mbox{sech}\left(\frac{x}{2 b_x \ell_{x,\mathrm{S}}}\right)\,,
\label{e_1dsechansatz}
\end{equation}
to describe the intermediate regime $\omega_x>0$. Substituting theses ansatz
into Eq.\ (\ref{e_1dhamiltonian}) we obtain an energy functional in terms of
the dimensionless length $\ell_x$. Note that the sech ansatz
length $\ell_{x,\mathrm{S}}$ is defined so that $\ell_{x,\mathrm{S}} \rightarrow 1$
as $\omega_x \rightarrow 0$, while the Gaussian ansatz length
$\ell_{x,\mathrm{G}}$ is defined such that $\ell_{x,\mathrm{G}} \rightarrow 1$ as
$\omega_x \rightarrow \infty$.

\begin{figure}[t]
\centering
\includegraphics[width=0.75\textwidth]{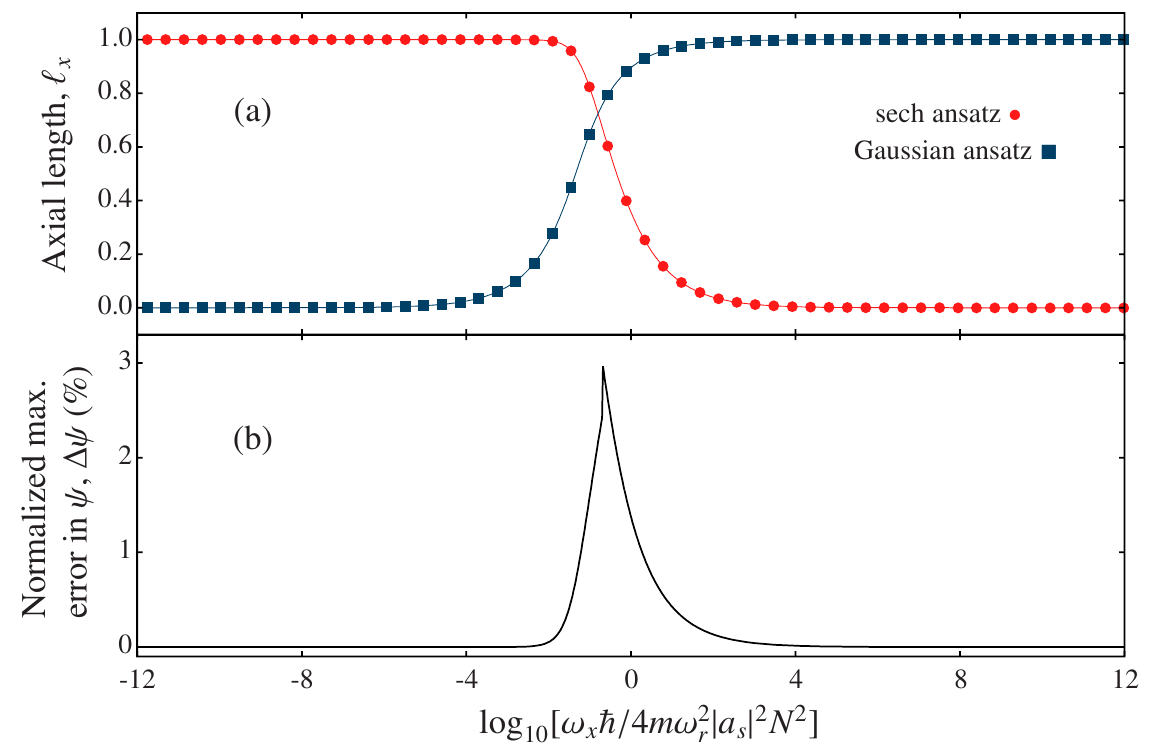}
\caption{\small Gaussian-ansatz [Eq.\ (\ref{e_1dgaussansatz})] and sech-ansatz [Eq.\
(\ref{e_1dsechansatz})] solutions of the 1D GPE [Eq.\ (\ref{e_q1dgpe})],
as found in Ref.\ \cite{billam_etal_variational_2011}. In (a), the
energy-minimizing axial length for each ansatz is shown, as a function of
$\omega_x$. In (b) the maximum absolute difference between the best-fitting
ansatz, $\psi_\mathrm{Ansatz}$, and the exact numerical solution, $\psi_0$, is
shown as a percentage of the peak value of $\psi_0$. This can be expressed
mathematically as $\Delta\psi = 100\mathrm{max}(|\psi_\mathrm{Ansatz} -
\psi_0|) / \mathrm{max}(\psi_0)$. Our deliberate definition of the ansatz
such that $\ell_{x,\mathrm{S}} \rightarrow 1$ as $\omega_x \rightarrow 0$ and
$\ell_{x,\mathrm{G}} \rightarrow 1$ as $\omega_x \rightarrow \infty$ results in
the potentially confusing trend that $\ell_{x,\mathrm{G}} \rightarrow 0$ as
$\omega_x \rightarrow 0$ despite the fact that the \textit{physical} length of
the Gaussian ansatz tends to a non-zero constant in this limit.}
\label{f_1d_variational}
\end{figure}

In the Gaussian case, one obtains the energy functional,
\begin{equation}
H_\mathrm{1D}(\ell_{x,\mathrm{G}}) =
\hbar \omega_x \left(
\frac{1}{4\ell_{x,\mathrm{G}}^2}
+\frac{\ell_{x,\mathrm{G}}^2}{4}
-\frac{a_x |a_s| N}{\sqrt{2\pi}a_r^2 \ell_{x,\mathrm{G}}}
\right)\,,
\label{e_1dgaussenergy}
\end{equation}
where $a_x = \sqrt{\hbar/m\omega_x}$ and $a_r = \sqrt{\hbar/m\omega_r}$ are the
axial and radial harmonic oscillator lengths. In the sech case, one instead
obtains,
\begin{equation}
H_\mathrm{1D}(\ell_{x,\mathrm{S}}) =
m \omega_r^2 a_s^2 N^2 \left(
\frac{1}{6\ell_{x,\mathrm{S}}^2}
-\frac{1}{3\ell_{x,\mathrm{S}}}
+\frac{\pi b_x^4 \ell_{x,\mathrm{S}}^2}{24 a_x^4}
\right)\,.
\label{e_1dsechenergy}
\end{equation}
Either of these energy functionals can be analytically (or numerically)
minimized to give the corresponding, variational-energy-minimizing, axial
length $\ell_x$ \cite{billam_etal_variational_2011}. The axial lengths $\ell_x$
for both variational solutions are shown in Fig.\ \ref{f_1d_variational}(a) as
a function of the axial trap frequency $\omega_x$.

\subsubsection{Comparison to bright soliton solution} The variational solutions
can be compared with a full numerical solution of
the 1D GPE [Eq.\ (\ref{e_q1dgpe})] to give an idea of how the axial
trapping affects the ground state \cite{billam_etal_variational_2011}. The
results of such an analysis are shown in Fig.\ \ref{f_1d_variational}(b), which
shows the maximum difference in shape between the lowest-energy (and hence,
most accurate) variational solution and the numerically exact ground state. As
one would expect, the sech ansatz converges to
the exact solution in the axially free limit $\omega_x\rightarrow0$ and it is
in this regime, where this ansatz approximates the exact solution well, that
the ground state can be regarded as soliton-like.  In the opposite,
trap-dominated limit $\omega_x \rightarrow \infty$, the Gaussian ansatz converges to the exact solution, which is no longer
soliton-like in appearance. Convergence is also somewhat slower for the
Gaussian ansatz as $\omega_x \rightarrow \infty$ than for the sech ansatz as
$\omega_x \rightarrow 0$ due to the density-dependent nature of the
nonlinearity \cite{billam_etal_variational_2011}.  In intermediate regimes, one
or other ansatz provides a good approximation to the solution over a wide range
of trap strengths, with only a small gap in which neither ansatz is
particularly accurate. Consequently, one can usefully think of the ground state
being deformed from sech-shaped to Gaussian-shaped as $\omega_x$ is increased.

However, the preceding analysis assumes the validity of the quasi-1D
approximation.  To obtain a complete picture of the ground state, and its
relationship to the NLSE bright soliton, a full treatment of the 3D GPE is
required. We undertake such a treatment in the next section (Section
\ref{s_static3d}).

\section{Bright solitary waves in 3D: static properties}
\label{s_static3d}

The 3D GPE [Eq.\ (\ref{e_3dgpe})] is non-integrable and does not support true
bright solitons. Nonetheless, bright solitary matter waves can be observed
\cite{khaykovich_etal_science_2002, strecker_etal_nature_2002,
cornish_etal_prl_2006} which continue to exhibit soliton-like behaviour for a
wide range of parameters. They are particularly soliton-like in their dynamical
properties --- especially when considering their mutual interactions and
collisions. In this section, however, we focus on the static regime by considering the solitary wave ground state solutions of the 3D GPE.

The existence and form of bright solitary wave ground states in the 3D GPE is
intricately linked to the instability of attractive condensates to collapse. In
this section we review the properties of stationary 3D bright solitary matter
waves in detail.  In Section \ref{s_collapse_parameter} we discuss the collapse phenomena.
In Section \ref{s_analysecollapse} we present variational and numerical
approaches to the problem. In Section \ref{s_ground_state_properties} we review
the properties of bright solitary waves as elucidated by the variational and
numerical methods, and compare the 3D results to the predictions of the
1D approach considered in Section \ref{s_static1d}.

\subsection{Collapse and the critical parameter}
\label{s_collapse_parameter}

An attractively-interacting
BEC in 3D is prone to a collapse instability.  Indeed, in the absence of trapping, the system will undergo collapse. Importantly, the presence of trapping can support metastable, non-collapsing states, although these existence of the metastable state depends on the atom number, interaction strength and shape and strength of the trapping potential. The collapse instability has been investigated experimentally \cite{donley_etal_nature_2001, roberts_etal_prl_2001,
gerton_etal_nature_2000, bradley_etal_prl_1997}.  Numerous theoretical studies have focused on identifying the
parameters associated with the onset of collapse in condensates of various geometries, using
variational \cite{carr_castin_pra_2002, billam_etal_variational_2011,
parker_etal_jpb_2007, perez-garcia_etal_pra_1997, salasnich_etal_pra_2002a},
perturbative \cite{yukalov_yukalova_pra_2005}, and numerical
\cite{carr_castin_pra_2002, gammal_etal_pra_2001, gammal_etal_pra_2002,
billam_etal_variational_2011, parker_etal_jpb_2007, ruprecht_etal_pra_1995,
dodd_etal_pra_1996} methods.   The condensate dynamics during collapse are the subject of continuing
theoretical study \cite{savage_etal_pra_2003, saito_ueda_prl_2001,
wuster_etal_pra_2005, dabrowska_wuster_etal_njp_2009, altin_etal_pra_2011}.

Recall, we parameterise the interaction strength of the condensate via $k = N |a_s|/ a_r$.
The relevance of $k$ is that, when it exceeds a critical value $k_c$, the metastable states cease to exist and the collapse phenomenon kicks in.  The value of $k_c$ is dependent on the trap geometry.

\subsection{Variational and numerical approaches to the static solutions}
\label{s_analysecollapse}

The parameter regime of metastable solutions of the 3D GPE with $a_s<0$
is most accurately determined by numerically solving the 3D GPE.   However, as shown in Section \ref{s_axialtrapping} in 1D, a variational approach can give insightful and accurate results.  Hence we begin with this approach using two variational ansatz: an ansatz with Gaussian radial and axial profiles, and an ansatz
with a Gaussian radial profile and a sech axial profile.

\subsubsection{Variational analysis: Gaussian ansatz}
\label{s_gaussianansatz}
\begin{figure}[t]
\centering
\includegraphics[width=11.7cm]{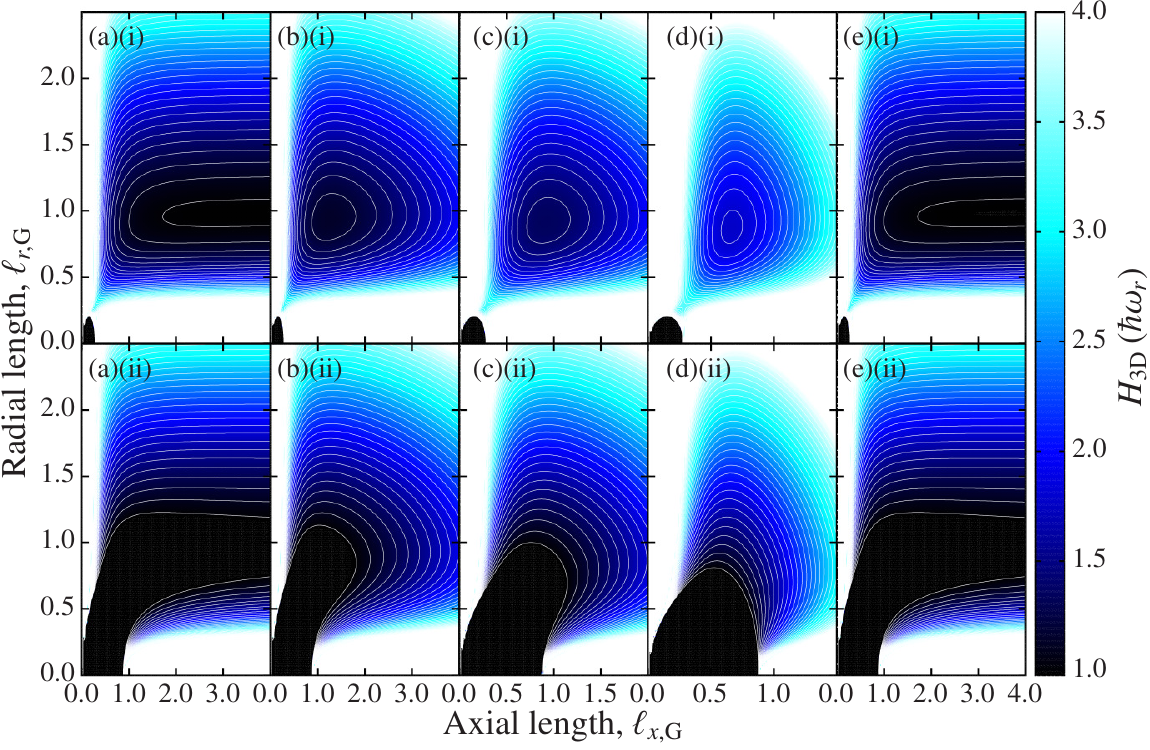}
\caption{\small Per-particle energy functional, $H_\mathrm{3D}$, determined using a
Gaussian ansatz [Eq.\ (\ref{e_gaussianansatz})] for a BEC in a cylindrically
symmetric, harmonic trap. Trap anisotropies shown are: (a)
$\lambda=0$, (b) $\lambda=1/2$, (c) $\lambda=1$, (d) $\lambda=2$, (e)
$\lambda^2=-4\times10^{-4}$ (expulsive axial potential). The top row [sub-label (i)] shows
the case $k=0.35$, for which all the trap geometries are stable to collapse. In
this case there is a stable local minimum in the variational
energy, which corresponds to the (metastable) bright solitary matter wave
ground state. The bottom row [sub-label (ii)] shows the case $k=1.1$, for which
all the trap geometries are unstable to collapse.}
\label{f_gauss_energy}
\end{figure}

The solution of the 3D GPE under cylindrically symmetric trapping can be
approximated by a normalized Gaussian ansatz of the form,
\begin{equation}
\left(\frac {1}{\pi^{3/2} a_r^3 \ell_{r,\mathrm{G}}^{2} \ell_{x,\mathrm{G}}}\right)^{1/2}
\exp\left(-\frac{1}{2a_r^2}\left[\frac{x^2}{\ell_{x,\mathrm{G}}^2}+\frac{r^2}{\ell_{r,\mathrm{G}}^2} \right] \right)\,,
\label{e_gaussianansatz}
\end{equation}
where $\ell_{x,\mathrm{G}}$ and $\ell_{r,\mathrm{G}}$ are, respectively, axial and radial variational
length parameters associated with the Gaussian ansatz. (Of course, this becomes the exact solution in the noninteracting regime ($a_{\rm s}=0$). ) Such an ansatz has been considered for bright solitary waves in
\cite{billam_etal_variational_2011, parker_etal_jpb_2007,
perez-garcia_etal_pra_1997}, and is most appropriate in parameter regimes where
the strength of the trap potential dominates over the strength of interactions
in all directions.  Substituting this Gaussian ansatz [Eq.\
(\ref{e_gaussianansatz})] into the classical field Hamiltonian for Eq.\
(\ref{e_3dgpe}),
\begin{equation}
H_\mathrm{3D}[\psi] = \int d\mathbf{r} \left[
\frac{\hbar^{2}}{2m}\left|\nabla\psi(\mathbf{r})\right|^{2}
+V(\mathbf{r})|\psi(\mathbf{r})|^{2}
-\frac{2\pi N|a_s|\hbar^2}{m}|\psi(\mathbf{r})|^{4}
\right]\,,
\label{e_3dhamiltonian}
\end{equation}
where $V(\mathbf{r})= m \omega_r^2 (\lambda^2 x^2 + r^2)/2$,
yields
\begin{equation}
H_\mathrm{3D}[\psi] =
\hbar \omega_r \left(
\frac{1}{4\ell_{x,\mathrm{G}}^2}
+\frac{1}{2\ell_{r,\mathrm{G}}^2}
+\frac{\lambda^2 \ell_{x,\mathrm{G}}^2}{4}
+\frac{\ell_{r,\mathrm{G}}^2}{2}
-\frac{k}{\sqrt{2\pi}\ell_{r,\mathrm{G}}^{2}\ell_{x,\mathrm{G}}}
\right)\,.
\label{e_3denergygauss}
\end{equation}
This defines an ``energy landscape" in terms of the variational lengthscales $\ell_{x,\mathrm{G}}$ and $\ell_{r,\mathrm{G}}$, in which the variational solution corresponds to an energy minimum.  Typical energy landscapes for this Gaussian
variational ansatz are shown in Fig.\ \ref{f_gauss_energy}.  We seek the lengthscales that minimize this variational energy.  Differentiating with respect to each of the lengthscale variables produces, respectively, two coupled conditions for the variational energy-minimizing lengths,
\begin{equation}
\lambda^2\ell_{x,\mathrm{G}}^4 + \frac{2k\ell_{x,\mathrm{G}}}{\sqrt{2\pi}\ell_{r,\mathrm{G}}^2} - 1 = 0\,,
\label{e_gauss_diff_lx}
\end{equation}
and
\begin{equation}
\ell_{r,\mathrm{G}}^4 + \frac{2k}{\sqrt{2\pi}\ell_{x,\mathrm{G}}} - 1 = 0 \,.
\label{e_gauss_diff_lr}
\end{equation}
In the case of prolate and oblate
trap potentials these equations can be solved via straightforward iterative
procedures \cite{billam_etal_variational_2011}, while for the axially free
case an analytic solution can be found \cite{carr_castin_pra_2002,
billam_etal_variational_2011}.

\subsubsection{Variational analysis: sech ansatz}
\label{s_sechansatz}
One can take the same variational approach but with a normalized sech
ansatz of the form
\begin{equation}
\left(\frac{1}{4\pi a_r^3 \ell_{r,\mathrm{S}}^2 \ell_{x,\mathrm{S}}}\right)^{1/2}
\mbox{sech} \left( \frac{x}{2a_r\ell_{x,\mathrm{S}}} \right)
\exp \left( -\frac {r^2}{2 a_r^2  \ell_{r,\mathrm{S}}^2} \right)
\,,
\label{e_sechansatz}
\end{equation}
where $\ell_{x,\mathrm{S}}$ and $\ell_{r,\mathrm{S}}$ are, respectively, axial and radial variational
length parameters. Such an ansatz has been considered in
\cite{billam_etal_variational_2011, parker_etal_jpb_2007,
carr_castin_pra_2002}, and is most appropriate in parameter regimes where the
strength of the \textit{radial} trap potential  dominates over the strength of
interactions, but the strength of interactions dominates over the strength of
the \textit{axial} trap potential.  Following the above procedure, the variational energy expression now becomes,
\begin{equation}
H_\mathrm{3D}[\psi] =
\hbar \omega_r \left(
\frac{1}{6\ell_{x,\mathrm{S}}^2}
+\frac{1}{2\ell_{r,\mathrm{S}}^2}
+\frac{\pi^2\lambda^2\ell_{x,\mathrm{S}}^2}{24}
+\frac{\ell_{r,\mathrm{S}}^2}{2}
-\frac {k}{3 \ell_{r,\mathrm{S}}^2 \ell_{x,\mathrm{S}}}
\right)\,.
\label{e_3denergysech}
\end{equation}
yielding the two conditions for the energy-minimizing lengthscales,
\begin{equation}
\lambda^2\ell_{x,\mathrm{S}}^4 + \frac{4 k \ell_{x,\mathrm{S}}}{\pi^2 \ell_{r,\mathrm{S}}^2} - \frac{4}{\pi^2} = 0\,,
\label{e_sech_diff_lx}
\end{equation}
and
\begin{equation}
\ell_{r,\mathrm{S}}^4 + \frac{2k}{3\ell_{x,\mathrm{S}}}-1=0 \,,
\label{e_sech_diff_lr}
\end{equation}
The sech ansatz yields variational
energy landscapes which are qualitatively very similar to those yielded by the Gaussian ansatz \cite{carr_castin_pra_2002, parker_etal_jpb_2007}.

\subsubsection{Numerical approaches}
\label{s_collapsenumerical}
A variational approach to the stability of bright solitary matter waves in 3D
yields considerable qualitative insight, particularly with regard to the
collapse phenomenon.  However, the approach is not particularly accurate in its
prediction of the critical parameter $k_c$; the imposition of a certain shape
on the wavefunction via the variational ansatz causes variational methods to
consistently over-estimate $k_c$.  Consequently, a great deal of work in the
field of attractively-interacting BECs has focused on accurately identifying
$k_c$, for various trap configurations, via numerical solution of the 3D GPE.  The main approaches to solving the GPE numerically are reviewed in Ref. \cite{minguzzi_etal_physrep_2004}.  As in Section \ref{s_static1d}, the
numerical and variational results can also be compared in order to investigate
how bright-soliton-like the bright solitary matter wave ground states are in
terms of their shape; such a comparison is, however, only meaningful in cases
which approach the quasi-1D limit \cite{billam_etal_variational_2011}.

\begin{figure}[t]

\centering
\includegraphics[width=10cm]{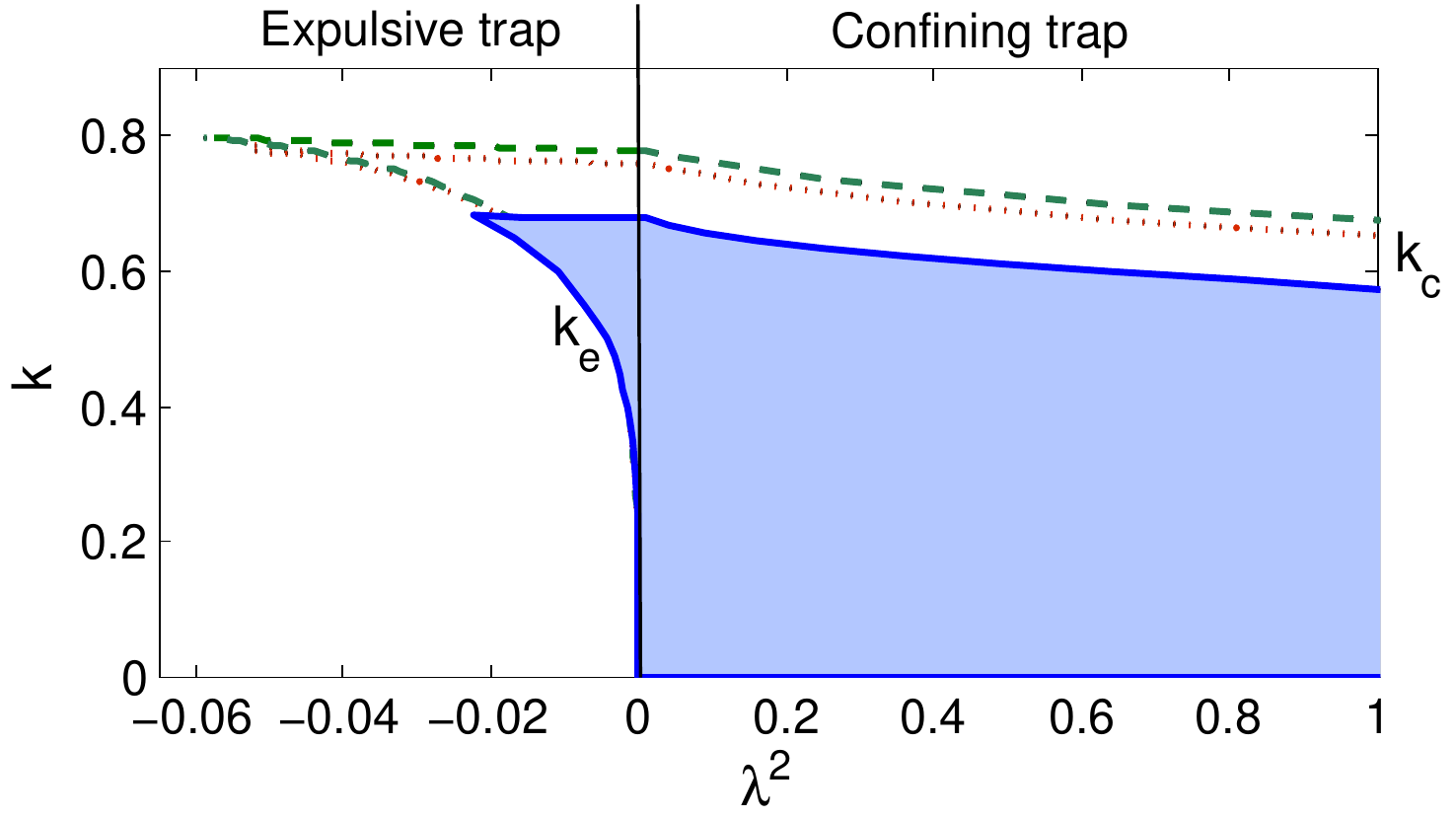}
\caption{\small Existence and properties of bright solitary matter wave ground states in cylindrically symmetric traps as a function of trap geometry, parameterized by $\lambda^2$.  The presence of metastable states is indicated by the various regions, according to the 3D GPE (blue/grey region), the Gaussian ansatz (region bound by the green dashed line) and the sech ansatz (region bound by the red dotted line).  The uppermost lines represent the critical parameter for collapse $k_c$; under an expulsive trap $\lambda^2<0$ there exists a lower bounding line representing the critical parameter for expansion $k_e$.  Note the difference in scale on the abscissa either side of $\lambda^2=0$ axis.  }
\label{f_cyl_solutions}       
\end{figure}

Studies have investigated traps with spherical \cite{ruprecht_etal_pra_1995,
dodd_etal_pra_1996} and cylindrical \cite{gammal_etal_pra_2001} symmetry,
cylindrically symmetric waveguides without axial trapping
\cite{parker_etal_jpb_2007}, and the case of a generally asymmetric trap
\cite{gammal_etal_pra_2002}. Several works also investigated the configurations
of specific experiments in detail \cite{parker_etal_jpb_2008,
carr_castin_pra_2002}.  The parameter space of bright solitary wave solutions,  under cylindrically-symmetric trapping, is summarized in Fig.~\ref{f_cyl_solutions}.

\subsection{Static solutions in 3D and the role of trapping}
\label{s_ground_state_properties}


Here we discuss the predicted bright solitary matter wave solutions (in cylindrically symmetric traps) according to the variational  method and the full numerical solution.
The
structure of the energy surfaces described by the Gaussian [Eq.\
(\ref{e_3denergygauss})] 
is
illustrated for a selection of trap geometries and interaction strengths in %
Fig.\ \ref{f_gauss_energy}.
The collapse instability is manifest as an unbounded decrease of $H_\mathrm{3D}$ as the
variational lengths $\ell_x$ and $\ell_r$ tend to zero.  In cases where a
bright solitary wave ground state exists (upper rows in figures) it is
stabilized against collapse by an energy barrier (forming a local energy minimum in the energy surface); in cases where such an energy
barrier is not present (lower rows in figures), no bright solitary matter wave
ground state exists. The parameter space of metastable ground state solutions as predicted by
the variational methods is compared to numerical solutions of the cylindrically
symmetric 3D GPE in Fig.\ \ref{f_cyl_solutions}.  In this plot we use the parameter $\lambda^2$ to specify the trap geometry; this is because that, as well as considering the conventional case of {\em confining} axial potentials ($\lambda^2>0$) we will also consider the case of {\em expulsive} axial potentials ($\lambda^2<0$).  We will discuss the results of these figures below by separately discussing four key trapping regimes (specified in terms of $\lambda^2$).

\subparagraph{Zero axial potential ($\lambda^2=0$)}
The case of a zero axial potential (which is equivalent, more generally, to any constant uniform potential in the axial direction), results in a waveguide-like trap.  It leads to some algebraic simplification in the variational equations and, in the case of the sech ansatz, the
variational energy-minimizing lengths $\ell_x$ and $\ell_r$ and the critical
parameter $k_c = 3^{-1/4}$ can be found analytically
\cite{carr_castin_pra_2002, billam_etal_variational_2011}.

More insight into the physical situation can be gleaned from the corresponding variational energy
surfaces, shown in Fig.\ \ref{f_gauss_energy}(a).  The energy surface
forms a relatively flat ``plain'' for larger $\ell_x$ and $\ell_r$, with
sharply rising ``ridges'' occurring when either length becomes small. However,
the (negative) interaction term in the energy functional leads to a distinct
``chute'' \cite{carr_castin_pra_2002} at the meeting point of these two ridges
(when both $\ell_x$ and $\ell_r$ are small).  For low $k$ a raised saddle point
separates the chute from the plain, thus forming the local energy minimum of the metastable solution; as $k$ increases this saddle lowers, until
at $k=k_c$ it disappears and the entire parameter regime of the plain becomes
unstable. For the sech ansatz, this transition at $k_c =
1/3^{1/4}\approx0.76$ \cite{carr_castin_pra_2002,
billam_etal_variational_2011}. For the Gaussian ansatz the critical value is
$k_c \approx 0.778$ \cite{perez-garcia_etal_pra_1997}. For comparison, the
non-polynomial Schr\"{o}dinger equation (an extended quasi-1D approach)
predicts $k_c = 2/3$, through a simpler calculation
\cite{salasnich_etal_pra_2002b}.   The true mean-field result, obtained by numerical solution of the 3D GPE, is $k_c=0.675$ \cite{parker_etal_jpb_2007}.

Within the regime of metastable solutions, the solitary wave lengthscales vary with the interaction strengths.  For $k=0$ the axial lengthscale is effectively infinite.  As $k$ is increased the axial lengthscale reduces monotonically, until the point of collapse.  The radial lengthscale stays close to the radial harmonic oscillator length $a_r=\sqrt{\hbar/m\omega_r}$ throughout.  Interestingly, the solution approaches being spherical as the collapse point is reached \cite{carr_castin_pra_2002}.

In regimes where a bright solitary wave ground state does exist, the energy of
the saddle point relative to that of the local energy minimum on the plain sets
an energy scale at which the bright solitary wave ground state will be unstable
to collapse when excited. Excitations with sufficient energy could allow the
condensate to overcome the barrier formed by the saddle point and lead to a
dynamical collapse in which $\ell_x$ decreases to zero
\cite{pitaevskii_pla_1996, carr_castin_pra_2002, parker_etal_jpb_2007}. A
second channel of instability also arises; because the lack of an axial trap
results in a finite-valued energy as $\ell_x \rightarrow \infty$, there exists
a ``dispersive channel'' in which excitations of the ground state above a
certain energy threshold can lead to dynamics where $\ell_x$ increases without
bound \cite{carr_castin_pra_2002, parker_etal_jpb_2007}.

In Ref.\ \cite{billam_etal_variational_2011} the 3D solitary wave ground state in
the waveguide-like trap was compared to the NLSE bright soliton. It was found in
Ref.\ \cite{billam_etal_variational_2011} that, while it is possible to reach a
highly soliton-like ground state in a waveguide-like trap, it lies in an
experimentally challenging regime. Nonetheless, as we review in Section
\ref{s_dynamic1d} and \ref{s_dynamic3d}, the dynamics of 3D bright solitary
waves can be highly soliton-like even when their static shape does not closely
resemble the NLSE soliton.

\subparagraph{Prolate and isotropic traps ($0<\lambda^2 \leq 1$)}
For $0< \lambda^2 < 1$ the trap is prolate, i.e. elongated in $x$, while for $\lambda =1$ it is
spherically symmetric.  In such cases the variational solutions must be obtained numerically
\cite{billam_etal_variational_2011}.

The energy landscape under these potentials [with examples shown in Fig.\ \ref{f_gauss_energy}(b) and (c)] is similar to that for the
waveguide trap $\lambda=0$ in and around the region of the collapse instability.  Indeed, the critical point for removal of the metastable state depends quite weakly on $\lambda$, as evident from Fig.\ \ref{f_cyl_solutions}.  The only qualitative difference introduced into the variational energy by axial trapping arises in the high-$\ell_x$ limit, where the potential energy of the trap
leads to an infinite total energy in the limit $\ell_x \rightarrow \infty$,
eliminating the dispersive channel altogether.

In Ref.\ \cite{billam_etal_variational_2011} the solitary wave ground state in
a prolate trap was also compared to the NLSE bright soliton and similarly to above, the achievement of a soliton-like ground state was found to be highly experimentally challenging.

\subparagraph{Oblate trap ($\lambda^2 >1$)}

Such a trapping geometry, in which $\omega_x > \omega_r$, is not typical for the study of bright solitary matter waves,
as in this geometry no clear analogy can be drawn with an integrable NLSE.

Nonetheless, when an oblate trap possesses a metastable ground state it is indeed a
solitary wave under the definition used by \cite{morgan_etal_pra_1997}.  These ground states have been
studied using the 3D GPE \cite{parker_etal_jpb_2007}, and 2D
reductions with 3D effects \cite{salasnich_etal_pra_2009}. The variational energy surface [Fig. \ref{f_gauss_energy}(d)] is similar to the prolate/isotropic case.

\subparagraph{Expulsive axial potential ($\lambda^2<0$)}

The self-trapped nature of bright solitary matter waves means they can
withstand being placed in a trap with a weakly expulsive harmonic axial
potential ($\lambda^2<0$) without dispersing. This was the case in the
experiment of Ref.\ \cite{khaykovich_etal_science_2002}, and considered theoretically in \cite{carr_castin_pra_2002,
parker_etal_jpb_2007}.

The ensuing variational energy surfaces, shown in Fig.\ \ref{f_gauss_energy}(e),
again (i) permit metastable states [Fig.\ \ref{f_gauss_energy}(e)i] and (ii) fully collapsed scenarios for $k > k_c$.   However,
the expulsive potential leads to a second instability via an ``expansive channel''
\cite{parker_etal_jpb_2007}.  This corresponds to axial spreading of
the solutions $\ell_x \rightarrow \infty$.  In contrast to the dispersive
channel --- which never completely prevents the existence of a metastable ground
state, but renders it unstable to (potentially very small) excitations --- the
expansive channel can destabilize the ground state; like the collapse channel's
``chute'', the expansive channel is separated from the ground state by a saddle
point, which disappears for sufficiently low $k$, or high $|\lambda|$. This introduces a
critical expansion parameter $k_e$, such that one must have $k_e < k <
k_c$ in order to observe a metastable ground state. The structure of $k_c$ and
$k_e$ is illustrated in Fig.\ \ref{f_cyl_solutions}; it is immediately apparent
that the regime of metastable ground state solutions with an expulsive axial
potential is severely restricted compared to the other cases. In particular,
$|\lambda|$ must be relatively close to zero to avoid passing the cusp point ($k_c=k_e$), beyond which metastable solutions are no longer found.

\subsubsection{Asymmetric trap potentials}
\label{s_asymmetric_3dstatic}

Removing the restriction to cylindrically symmetric trap geometries which we
have enforced up to now leads to a considerably enlarged parameter space to
explore. The critical parameter in such traps has been
numerically determined by Gammal \textit{et al.}
\cite{gammal_etal_pra_2002}.  The existence and form of the bright solitary wave ground
state in anisotropic traps shows no qualitative differences from the
cylindrically symmetric case.


\section{Bright solitary waves in 1D: dynamics}
\label{s_dynamic1d}

When analysing the dynamics of bright solitary matter waves, it is naturally of
interest to compare their dynamics to the well-known and rich dynamics of
bright solitons. The natural regime for such comparison is the case in which
the axial potential is weak compared to the radial trap potential. As discussed
in Sections \ref{s_static1d} and \ref{s_static3d}, this regime is where
stationary bright solitary matter waves bear the greatest resemblance to bright
solitons, and experiments to date have focused on this regime.

Under a quasi-1D geometry, the condensate dynamics are
described by the 1D GPE [Eq.\ (\ref{e_q1dgpe})].  In the case of zero
axial trapping this reduces further to the focusing NLSE, admitting exact
bright solitons.  With a weak trapping or expulsive axial potential, as
realized in experiments, integrability is lost and the dynamics are no longer
those of true solitons. Nonetheless, as we illustrate in this section, the
dynamics remain highly soliton-like under the assumption that a 1D
description is accurate. We shall later relax this 1D assumption in
Section \ref{s_dynamic3d}.

In this section, we begin by reviewing the dynamics of single and multiple NLSE
bright solitons (Section \ref{s_nlse_dynamics}), and introduce a particle-like
model for their motion and interactions. Introducing axial trapping, we then
explore the dynamics of bright solitary waves in the quasi-1D approximation
(Section \ref{s_quasi1d_dynamics}); these dynamics are highly soliton-like and
can be easily understood using a straightforward modification of the particle
model.

\subsection{Dynamics and collisions of the classic bright soliton}
\label{s_nlse_dynamics}
In the absence of axial trapping, an attractively-interacting BEC in a quasi-1D
trap is described by the focusing 1D NLSE, which supports bright soliton
solutions \cite{zakharov_shabat_1972, zakharov_shabat_1972_russian}. The
single- and many-soliton solutions to this equation have been extensively
explored in the context of optical solitons \cite{zakharov_shabat_1972,
zakharov_shabat_1972_russian, satsuma_yajima_1974, gordon_ol_1983,
kodama_hasegawa_1991, afanasjev_vysloukh_j_opt_soc_am_b_1994}. We review the
results pertinent to soliton dynamics in this section.

\subsubsection{Dynamic bright soliton solutions}
\label{s_nlse_soliton_solutions}
Eliminating the axial trapping in the 1D GPE [Eq. (\ref{e_q1dgpe})]
yields the focusing NLSE,
\begin{equation}
i \hbar \frac{\partial \psi(x,t)}{\partial t} =
-\frac{\hbar^2}{2m}\frac{\partial^2 \psi(x,t)}{\partial x^2}
- 2 \hbar \omega_r  N |a_s| |\psi(x,t)|^2 \psi(x,t)\,.
\label{e_nlse}
\end{equation}
Despite its nonlinear nature, the integrability of Eq.\ (\ref{e_nlse}) means
solutions can be found using the inverse scattering method
\cite{zakharov_shabat_1972, zakharov_shabat_1972_russian}. In summary, a
scattering transform of $\psi(x,t)$ yields, at any time $t$, a spectral
decomposition of $\psi(x,t)$ into solitons and radiation. The radiation part of
the spectrum is continuous, and has in general a non-trivial time-dependence.
However, the soliton part of the spectrum is discrete and
\textit{time-independent}, and is completely described by four real quantities
for each soliton. Consequently, the spectrum of an $N$-soliton solution with no
radiation component can be completely described by $4N$ real quantities, from
which the complete solution $\psi(x,t)$ can be recovered using the inverse
scattering transform.

The most general $N$-soliton solution to Eq.\ (\ref{e_nlse}), containing no
radiation, can be written as
\cite{gordon_ol_1983},
\begin{equation}
\psi(x,t) = \sum_{j=1}^N \psi_j(x,t)\,,
\label{e_solitonsum}
\end{equation}
where,
\begin{equation}
\sum_{j=1}^N \frac{\gamma_i^{-1}+\gamma_j^\ast}
{\lambda_i + \lambda_j^\ast}
\psi_j(x,t)
= \frac{1}{\sqrt{b_x}}\,.
\label{e_solitonlineqn}
\end{equation}
Here we have defined the quantities,
\begin{equation}
\lambda_j = \frac{a_j}{2} + \frac{iv_j}{2\omega_r|a_s|N}\,,
\end{equation}
and,
\begin{equation}
\gamma_j = \exp \left[
\lambda_j \left( \frac{x-x_{j}}{b_x} \right)
+i \lambda_j^2 \frac{2 m \omega_r^2 a_s^2 N^2}{\hbar}t
+i\Phi_{j} \right]\,,
\end{equation}
in addition to the characteristic soliton length $b_x = \hbar / 2 m \omega_r
|a_s| N$. Each soliton is described by a real amplitude $a_j$, velocity $v_j$,
position offset $x_{j}$, and phase $\Phi_{j}$. In the case that the $j$th
soliton is well-separated from the other $N-1$ solitons, the linear system
defined by Eq.\ (\ref{e_solitonlineqn}) can be approximately solved to give
\cite{gordon_ol_1983},
\begin{eqnarray}
\label{e_soliton}
\psi(x,t) &=& \frac{a_j}{2\sqrt{b_x}}
\mbox{sech}\left(\frac{a_j(x-x_j-v_j t)}{2b_x} + q_j\right)\\
& & \mbox{} \times\exp\left(
i\left\{
\frac{m}{\hbar} \left(v_j(x-x_j)
+ \frac{v_j^2 t}{2}
+ \frac{\omega_r^2 a_s^2 N^2 a_j^2 t}{2} \right)
+ \Phi_j + \Psi_j
\right\} \right)\,.\nonumber
\end{eqnarray}
Here, $q_j$ and $\Psi_j$ are time-dependent position- and phase-shifts which
appear as a result of collisions with the other $N-1$ solitons. They are given
by,
\begin{equation}
q_j + i \Psi_j = \sum_{k \ne j} \pm \log
\left(\frac{a_j + a_k + i(v_j-v_k)/2\omega_r |a_s| N}
{a_j - a_k + i(v_j-v_k)/2\omega_r |a_s| N} \right)\,,
\end{equation}
where the sign is positive (negative) when the $j$th soliton is to the left (right) of the
$k$th \cite{gordon_ol_1983}.  While the $j$th soliton is
well-separated these shifts remain approximately constant, and only change
significantly during collisions.

\subsubsection{Bright soliton dynamics and collisions}
\label{s_nlse_soliton_dynamics}

The dynamics of a single bright soliton
in the NLSE are determined entirely by their nonlinear interactions with the
remainder of the solution. It is convenient to divide the remainder of the
solution into soliton and radiation components, and consider the influence of
these components on the dynamics separately.  We review soliton dynamics due to
soliton interactions in this section. The majority of these dynamics can be
understood on the basis of a simple particle model. The interaction of solitons
with radiation is more mathematically involved \cite{satsuma_yajima_1974,
boffetta_osborne_jcp_1992, afanasjev_vysloukh_j_opt_soc_am_b_1994} and no
similarly general picture is available.

In the absence of radiation, the dynamics of multiple bright solitons are
dominated by the interactions and collisions between solitons. One of the
defining characteristics of true solitons, associated with the integrability of
the system, is that they survive mutual collisions entirely unchanged in form.
The only observable effects of the collision are the asymptotic position and
phase shifts discussed in Section \ref{s_nlse_soliton_solutions}.

\begin{figure}[t]
\centering
\includegraphics[width=11.7cm]{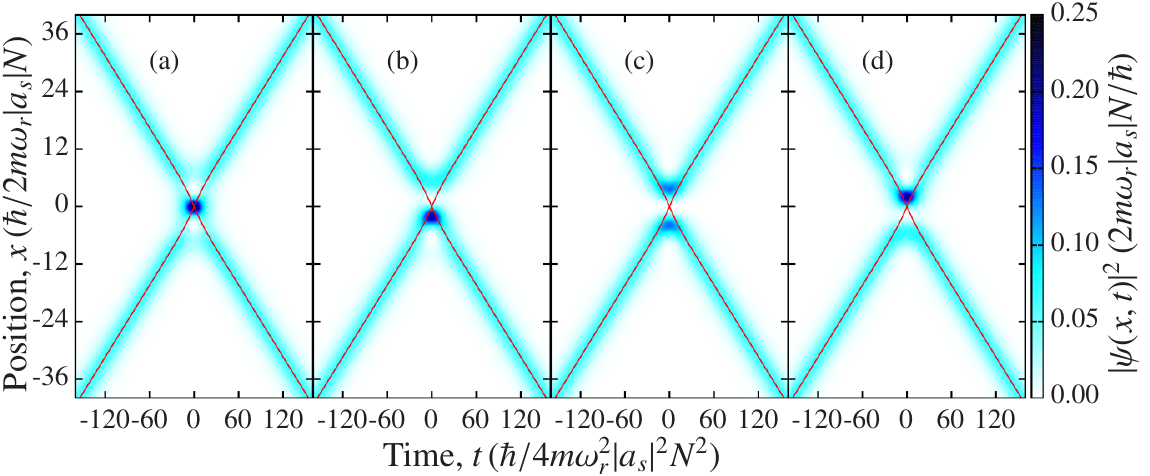}
\caption{\small Bright soliton collisions in the nonlinear Schr\"{o}dinger equation,
for solitons with equal amplitude and relative phase $\Delta\Phi =0$ (a),
$\pi/2$ (b), $\pi$ (c), and $3\pi/2$ (d). In each case the density profile of
the solution is superimposed with the soliton trajectories predicted by a
particle model \cite{martin_etal_prl_2007,martin_etal_pra_2008}. This
phase-independent model fails to describe the dynamics of the collision in
detail, but correctly incorporates the asymptotic position shift of the
solitons.}
\label{f_nlse_collisions}
\end{figure}

The main
characteristics of soliton interactions can be illustrated by the collisions of two
equal-amplitude solitons. This is shown, for various
relative phases $\Delta\Phi = \Phi_1-\Phi_2$,  in Fig.\
\ref{f_nlse_collisions}. As expected, the solitons survive such a collision
completely unchanged in form. The position shifts $q_j$ are clearly visible as
the deviation of both solitons from their initial linear trajectories. Although
the dynamics of the collision itself differ with the relative phase
$\Delta\Phi$, the position shift $q_j$ is unchanged.  Note that, due to the phase symmetry of the collision, the $0$-relative phase case leads to a central density anti-node, whereas for $\pi$-relative phase a density node is preserved at the origin.

The independence of the position shifts $q_j$ from the solitons' relative
phase $\Delta\Phi$ allows one, in principle, to predict their asymptotic
trajectories independently of their phase. Disregarding the phase information
in this way leaves each soliton described by a position, velocity, and
amplitude. One can then treat the solitons as classical particles with
an effective mass proportional to their amplitude and some appropriate
inter-particle potential. This approach was developed for optical NLSE solitons
\cite{maki_kodama_1986, scharf_bishop_pra_1992, scharf_bishop_pre_1993,
scharf_csf_1995}, using the inter-particle potential,
\begin{equation}
V(x_j-x_k) = - 2 \eta_j \eta_k (\eta_j+\eta_k)
\mbox{sech}^2 \left( \frac{2 \eta_j \eta_k (x_j-x_k)}{b_x (\eta_j + \eta_k)} \right)\,,
\end{equation}
where the solitons are treated as classical particles of effective mass $\eta_j
= a_j/4$. This potential reproduces the correct asymptotic position shifts
provided the velocities and effective masses satisfy the condition $|\eta_j -
\eta_k| \ll |v_j-v_k|/4\omega_r|a_s|N$ \cite{martin_etal_pra_2008}. The
particle model therefore reproduces the asymptotic shift exactly for the
equal-effective-mass solitons in Fig.\ \ref{f_nlse_collisions}.  For collisions of solitons with non-equal effective mass, the asymptotic shift predicted by the particle model approaches the correct value for weak soliton interactions (small density or $s$-wave scattering length) or short interaction times (high-velocity collisions).

\subsection{Dynamics and collisions under axial trapping}
\label{s_quasi1d_dynamics_collisions}

Despite the lack of integrability in the 1D GPE with an inhomogeneous axial
potential, and the resulting absence of true solitons, one may still observe
solitary waves if the stationary, eigenstate solutions of the equation can
propagate without changing shape \cite{morgan_etal_pra_1997}. While they do not
satisfy the strict mathematical requirements to be solitons \cite{fadeev,
ablowitz_segur_book}, these non-dispersive solitary waves can nonetheless,
under certain conditions, behave and interact in a soliton-like way.

\subsubsection{Dynamic bright solitary matter wave solutions}
\label{s_quasi1d_dynamics}

The possibility to observe solitary waves of this type was examined in considerable generality in Ref. \cite{morgan_etal_pra_1997}. In this work the authors considered, in 1, 2 and 3D, how static eigenstate solutions of nonlinear Schr\"{o}dinger equations with a general nonlinearity and an arbitrary external potential behaved when used as initial conditions in a nonlinear Schrodinger equation with the same nonlinearity and a new, possibly time-dependent, external potential. Two conditions were found to be necessary for the original eigenstates behave as solitary waves under the influence of the new potential: firstly, the nonlinearity must be decoupled from the absolute position, a requirement immediately satisfied by the conventional cubic form of the nonlinearity appearing in the GPE. Secondly, the new potential must differ no more than linearly in any spatial coordinate from the original potential \cite{morgan_etal_pra_1997}. We note that this second condition implies that one can use a time-dependent linear potential as a way to control bright solitary waves in experiments without causing them to lose their solitary-wave character; experimental control techniques such as this are discussed further in Section \ref{s_outlook}.

We shall restrict our attention to the case where the eigenstates are the bright solitary wave ground states considered in Section 2, \footnote{It is also possible to consider solitary waves having
the form of higher-energy nonlinear eigenstates; such eigenstates were
considered in Ref.  \cite{kivshar_etal_pla_2001}} and the ``new'' potential is a harmonic one of identical frequency to the original, but with a displaced center; this is equivalent to the case of the original potential acting on a displaced bright solitary wave ground state. In this case, the solitary wave has the same spatial profile as the ground state, but its centre of mass moves as a classical
particle in the static harmonic potential. If free from the influence of other
solitary waves or other components of the solution, it thus undergoes simple
harmonic motion like a classical particle \cite{morgan_etal_pra_1997,
martin_etal_prl_2007, martin_etal_pra_2008}. This feature of the GPE ground state is analogous to the Kohn theorem for the many-body ground
state. The latter guarantees that the true quantum mechanical ground state of
$N$ bosons in a harmonic trap can be expressed as a separable tensor product of
a single-body wavefunction in the centre of mass coordinate with a general
$(N-1)$-body wavefunction in the remaining inter-particle coordinates.

\subsubsection{Bright solitary matter wave collisions under axial trapping}
\label{s_quasi1d_collisions}

If the bright solitary wave is not well-separated from other components of the solution, its dynamics are influenced by the nonlinear
interaction with the other components.  As in the case of bright solitons, we
concentrate on the interactions between solitary wave components only.  For bright solitons the asymptotic effects of
the interactions were entirely described by phase and position shifts (Section
\ref{s_nlse_soliton_dynamics}). For bright solitary waves this is no longer
strictly true; however, this can be considered a satisfactory approximation in
the limit that the external potential is approximately constant over the region
of the collision and provided that the solitary waves are approximately
bright-soliton-shaped.  Making these approximations, one can combine the
particle model of soliton collisions (Section \ref{s_nlse_soliton_dynamics})
with the behaviour of a particle in a harmonic trap \cite{martin_etal_prl_2007,
martin_etal_pra_2008}. This leads to a combined particle model for multiple
bright solitary waves in a harmonic trap, which is most accurate for (a)
weak harmonic traps, (b) fast solitary wave collisions, and (c) in-phase
solitary wave collisions.

\begin{figure}[t]
\centering
\includegraphics[width=11.3cm]{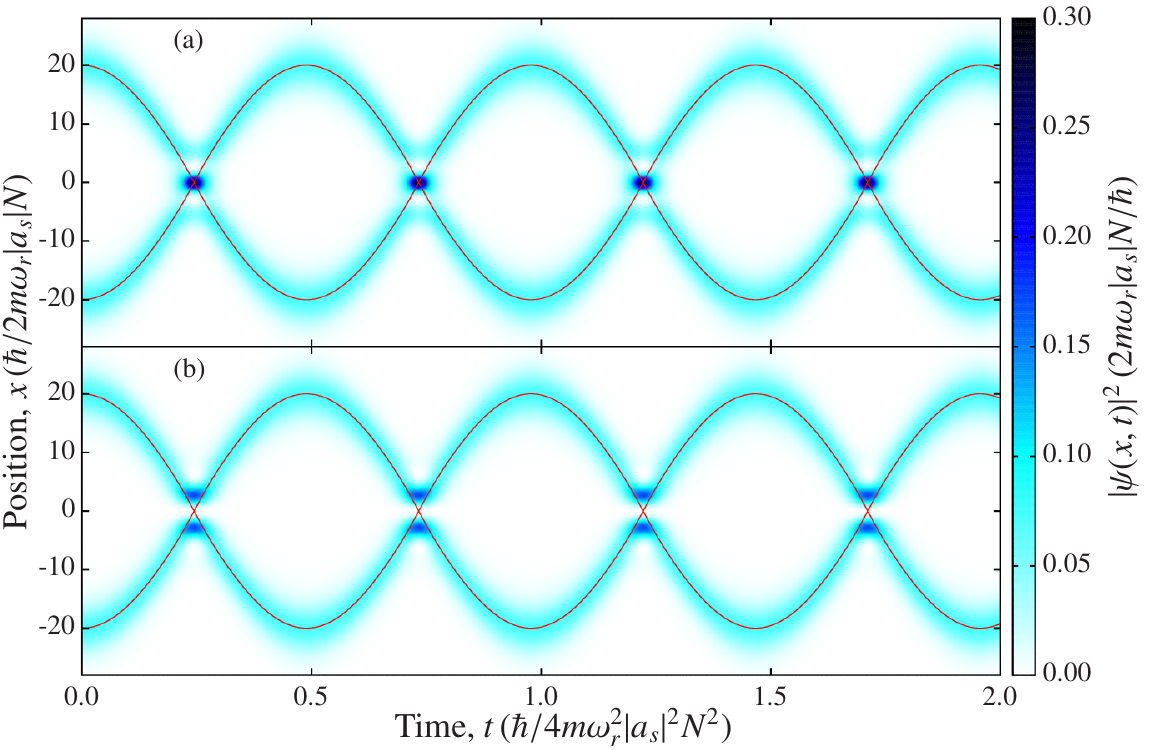}
\caption{\small Bright solitary wave collisions in the 1D GPE with harmonic axial trapping. The initial solitary waves, each with $N/2$ atoms, are
ground states of the trap displaced by $\pm\approx10.35$ in $x$ and
with zero initial velocity and relative phase $\Delta\Phi =0$ (a), and $\pi$
(b). In each case the density profile of the solution is superimposed with the
soliton trajectories predicted by the particle model.  }
\label{f_1d_collisions}
\end{figure}

The collisional dynamics of two, identical bright solitary waves according to the 1D GPE are
illustrated in Fig.\ \ref{f_1d_collisions} for relative phases of $0$ and $\pi$.  As
anticipated by the particle model, the dynamics are dominated by harmonic
particle-like motion when the waves are well-separated; however, when the waves
collide, periodically, at the trap centre, a soliton-like collision results in
a position shift. There is no overall phase shift between collisions, however
\cite{martin_etal_pra_2008}.  During the collision, we obtain qualitatively the same phase-dependent density structure as for the bright soliton collisions. The particle model predicts these trajectories
well over the short times shown here. However, over longer times deviations
do build up, arising from the variation of the
harmonic axial potential over the characteristic length scale of the collision \cite{martin_etal_prl_2007,
martin_etal_pra_2008}.

The complex dynamics of three or more solitary
waves oscillating and colliding in a harmonic trap can be effectively predicted
using the particle model; interestingly, the model is itself non-integrable for
three or more solitary waves, leading to chaotic particle-like dynamics
\cite{martin_etal_prl_2007, martin_etal_pra_2008}.

\section{Bright solitary waves in 3D: dynamics}
\label{s_dynamic3d}

The solitary waves of the 1D GPE explored in the previous section are in
many respects similar to NLSE bright solitons despite the addition of a trap
potential. However, in real experiments it is not only the addition of trapping
which leads to deviation from the NLSE, but also three-dimensional effects.
While the 3D ground state is still a solitary wave \cite{morgan_etal_pra_1997},
residual 3D effects can lead to large deviations from soliton-like behaviour,
although there are regimes where highly soliton-like dynamics can still be
observed.

Although the absence of an axial trap potential does not lead to exact soliton
solutions in 3D, we nonetheless begin by considering the axially free
case in Section \ref{s_dynamic3d_waveguide}. We then consider the additional
effects of an axial trap in Section \ref{s_dynamic3d_trap}. In broad parallel
to the previous section, we focus on the dynamics and interactions of solitary
waves only.  However, this distinction is blurred due to the 3D effects during collisions,
which can lead to non-soliton-like behaviour and eventual destruction of
solitary waves. This behaviour is fundamentally linked to the collapse
instability in 3D, explored in Section \ref{s_static3d}.

\subsection{Dynamics and collisions in a waveguide}
\label{s_dynamic3d_waveguide}

In this section we consider a waveguide-like trap, with harmonic radial and
zero axial trapping potential.  In such a trap the solitary wave profile is the
ground state, in the parameter regimes that are stable against
collapse (Section \ref{s_static3d}). With uniform axial potential this solitary wave is
self-trapped in the $x$-direction, and the dynamics of multiple such solitary
waves is consequently dominated by their interactions, as for NLSE bright
solitons.

\subsubsection{Stability of solitary wave collisions}
\label{s_dynamic3d_waveguide_stability}

In the absence of analytic solutions for binary solitary wave collisions in a
waveguide trap, such collisions must be simulated numerically. This can be done
from an initial condition composed of two copies of the (numerically obtained)
ground state, displaced from each other by some distance
and given some velocity toward each other\footnote{Such a velocity is imparted
numerically by applying a spatially varying phase of $e^{\pm i m v x/\hbar}$
Experimentally, this could be achieved by applying a linear external potential
to each solitary wave for a short time.}.  For equal-sized solitary
waves the resulting collisions can be studied within the parameter space of
incident velocity $v$, interaction strength parameter $k$, and relative phase
$\Delta\Phi$ \cite{parker_etal_jpb_2008}.

\begin{figure}[t]
\centering
\includegraphics[width=5.15in]{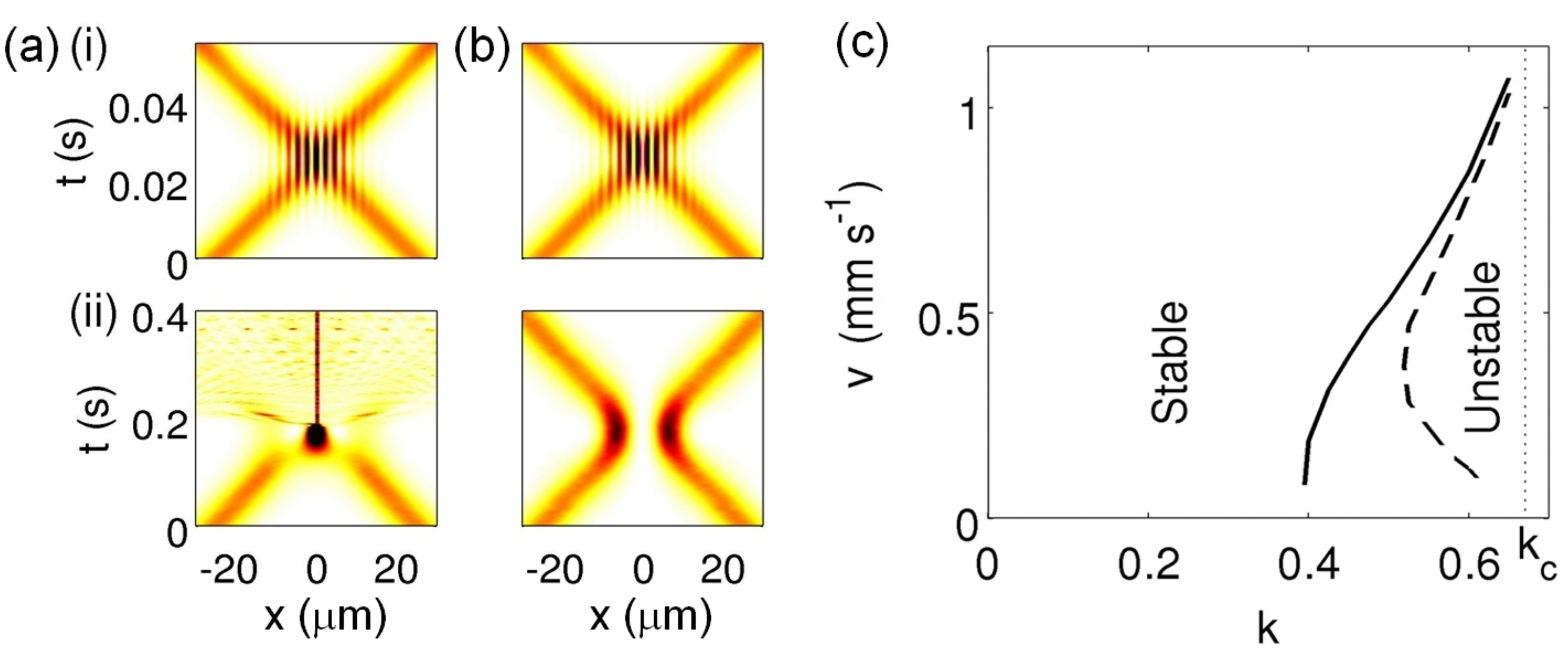}
\caption{\small Stability of solitary wave collisions in an axially homogeneous waveguide potential.  (a) Space-time plots of atom density during collision between two identical solitary waves, each with $k=0.4$ and featuring $\Delta \Phi =0$, for (i) high incident speed and (ii) low incident speed.  (b) The same but for $\Delta \Phi = \pi$.  (c) Stability diagram of the solitary wave collisions as a function of incident speed $v_i$ and interaction parameter $k$, where the solid (dashed) line marks the boundary between stable and unstable collisions for $\Delta \Phi = 0~(\pi)$ .  The vertical line denoted $k_c$ indicates the critical interaction strength for collapse of a single, isolated bright solitary wave.  The figure and parameters are from \cite{parker_etal_jpb_2008}.}
\label{f_3d_collisions_homogeneous}
\end{figure}

As for the ground state itself, the key parameter determining the stability of
collisions of this type is the interaction strength parameter $k$; this must remain below
some threshold $k_\mathrm{c}$ in order to avoid a dynamically-induced
collapse when the waves meet. However, $k_\mathrm{c}$ itself is dependent on the other
collision parameters. In particular, $k_\mathrm{c}$ is larger for faster
collisions, and for collisions with a relative phase closer to $\pi$. This is evident from the simulated results in Fig. \ref{f_3d_collisions_homogeneous}(a) and (b).  The
latter effect is most noticeable for low velocities, with the phase-dependence
of $k_\mathrm{c}$ disappearing in the high-velocity limit.\footnote{The GPE is based on the assumption of atomic scattering at low energy and momentum and so by ``high velocity" here we refer to a scale relative to the condensate's natural speed scale of the speed of sound $c=\sqrt{4 \pi \hbar a_s n/m^2}$\cite{pethick_smith_2002}} At low velocity, this
phase-dependence can be understood from the collision profiles illustrated for
the NLSE in Fig.\ \ref{f_nlse_collisions}; in the case $\Delta\Phi=\pi$ the
density profile of the collision itself resembles two solitons interacting
\textit{repulsively} \cite{gordon_ol_1983} and never overlapping, whereas
in the case $\Delta\Phi =0$ the solitons overlap, leading to a strong density
peak. While this peak is of no consequence in the NLSE or the 1D GPE, in
the 3D GPE this peak in the atomic density can trigger the collapse
instability.  The full dependence of the collisional stability on $k$ and incident speed $v$ is shown in Fig.~\ref{f_3d_collisions_homogeneous} for the cases of $\Delta \Phi=0$ and $\Delta \Phi=\pi$.  Note that the phase dependence of the collisional stability is also
predicted by effective 1D equations retaining more 3D character than the
1D GPE \cite{khaykovich_malomed_pra_2006}.

The dependence of $k_\mathrm{c}$ on the incident velocity $v$ can be
understood in terms of the relationship between the characteristic time for
collapse of the condensate, $t_\mathrm{c}$, and the characteristic time
for the collision-interaction to take place, $t_\mathrm{int}$. In Ref.
\cite{parker_etal_jpb_2008} it was illustrated that the critical collision
velocity, below which collapse occurred in numerical simulations of collisions,
for the parameters of the JILA solitary wave experiment
\cite{cornish_etal_prl_2006}, corresponds to a collision-interaction time
$t_\mathrm{int}$ approximately equal to the experimentally measured collapse
time, $t_\mathrm{c}$. Theoretical investigation of the role of the two
timescales has not proceeded further to date, in part because the GPE has not
been generally considered an accurate predictor of $t_\mathrm{c}$.
However, recent results suggesting that the GPE can accurately predict
$t_\mathrm{c}$ when a three-body loss term
is included \cite{altin_etal_pra_2011} offer the possibility of further
progress in this area.

\subsubsection{Population transfer in solitary wave collisions}
\label{s_dynamic3d_waveguide_population}

\begin{figure}[t]
\centering
\includegraphics[width=4in]{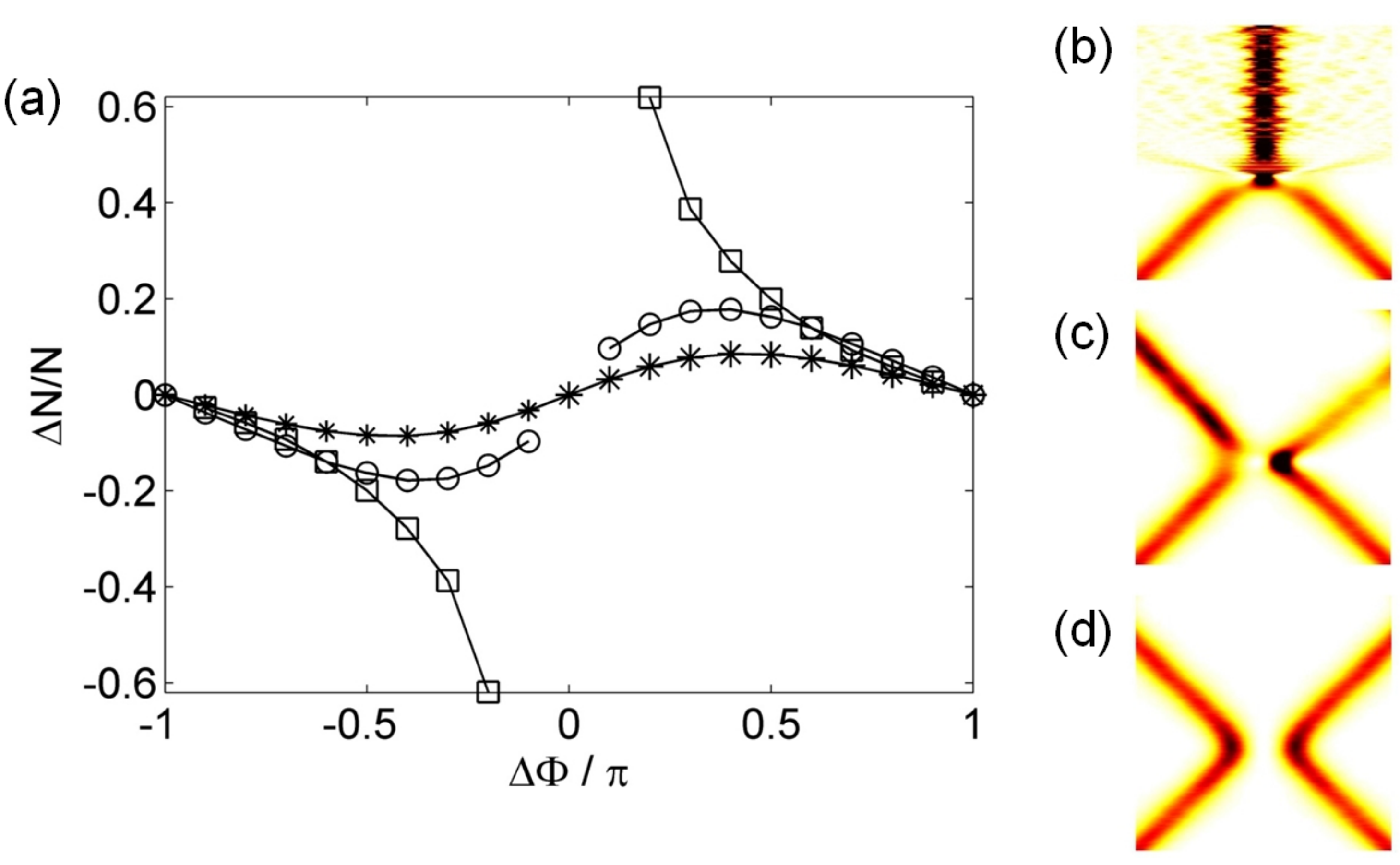}
\caption{\small (a) Population transfer $\Delta N$ during bright solitary wave collisions within a homogeneous waveguide, as a function of the relative phase between the waves $\Delta \Phi$.  At high incident speed (stars) the transfer is sinusoidal.  For reduced speed (circles) the transfer becomes larger and skewed towards $\Delta \Phi =0$.  For low speed (squares) the population transfer diverges as $\Delta \Phi \rightarrow 0$, due to runaway nonlinear effects and collapse.  The results are taken from \cite{parker_etal_jpb_2008}. (b-d) Wave dynamics for the low speed case with (b) $\Delta \Phi =0$, (c) $\pi/2$ and (d) $\pi$. }
\label{f_population_transfer}
\end{figure}

Another effect occurring as a result of the 3D nature of the system is that of
population transfer between bright solitary waves. In both the 1D NLSE, and the
3D GPE for a waveguide trap, collisions between solitons or solitary waves with
relative phases $\Delta\Phi =0 $ and $\pi$ have a density profile which remains
completely symmetric in $x$ after the collision; in this respect the 1D NLSE
and 3D GPE are analogous. The two descriptions, however, lead to very different dynamics
for intermediate phases $0<\Delta\Phi<\pi$ and $\pi < \Delta\Phi <2\pi$. In the 1D NLSE the density profile, which is initially symmetric in
$x$, loses its symmetry during the collision and regains it afterwards. In the
3D GPE for a waveguide trap, the initially-symmetric density profile loses its
symmetry during the collision, and this loss of symmetry leads to population
transfer between the two waves: the first solitary wave grows in amplitude and
slows down, while the second wave loses amplitude and speeds up.  Example dynamics are shown in Fig.\ \ref{f_population_transfer}(b-d).  In addition to
the 3D GPE, this effect can also be seen in effective-1D approaches
retaining extra 3D character \cite{khaykovich_malomed_pra_2006}.

The amount of population transferred shows
interesting dependencies on the relative phase and velocity of the solitary
waves (Fig.\ \ref{f_population_transfer}) \cite{parker_etal_jpb_2008}.  For fast collisions the amount of population transfer depends
sinusoidally on the relative phase, with the maximum transfer occurring at
$\Delta\Phi=\pi/2$ and $\Delta\Phi = 3\pi/2$, and the magnitude of this
transfer decreasing with velocity. At lower velocities, however, this dependence
becomes heavily skewed, with the maximum transfer occurring closer to $\Delta\Phi=0$ and the collapse instability occurring in extreme cases.  This deviation from sinusoidal transfer is a result of strong transient nonlinear effects during the collision
\cite{parker_etal_jpb_2008}.

\subsection{Dynamics and collisions under axial trapping}
\label{s_dynamic3d_trap}

We now introduce an axial harmonic trap, in addition to the waveguide-like trap
of the previous sections. In analogy to transition from the NLSE to the
1D GPE, such a system supports solitary wave excitations which have the
shape of the system ground state and move like classical particles in the
axial harmonic trap when well-separated from other solitary waves.  When these
solitary waves do interact and collide, two key questions arise: (a) how
soliton-like, and (b) how stable are these collisions? Both these questions are
particularly pertinent to the interpretation of bright solitary wave
experiments to date, and to unresolved issues regarding the role of relative
phase (as will be discussed in Section \ref{s_outlook}). We address each question in turn in the
following sections.

\subsubsection{Soliton-like dynamics}
\label{s_dynamic3d_trap_solitonlike}

\begin{figure}[t]
\centering
\includegraphics[width=3.8in]{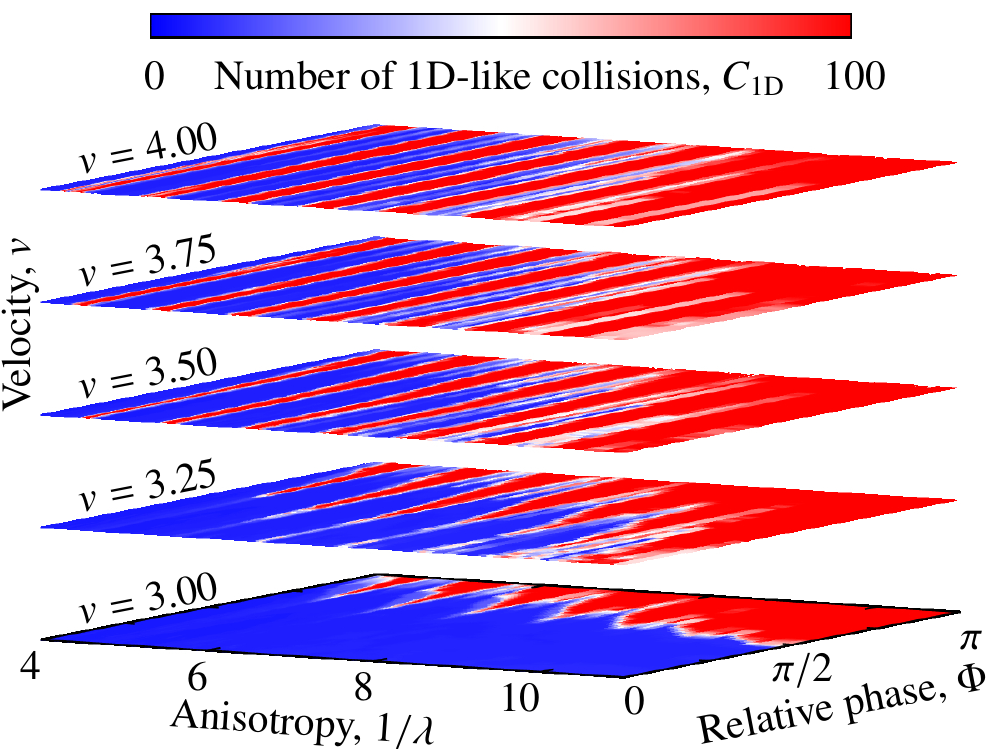}
\caption{\small Stability of bright solitary wave collisions in a cylindrically
symmetric 3D trap, as established in Ref.\ \cite{billam_etal_pra_2011}. The
number of 1D-like bright solitary wave collisions, $C_\mathrm{1D}$, is shown as
a function of the incident velocity $v$, relative phase $\Phi$, and trap
anisotropy $\lambda$. Here, $C_\mathrm{1D}$ is defined as the number of
collisions for which the solitary waves return to within 75\% of their original
peak amplitude and position at their maximum distance from the origin. Higher
$C_\mathrm{1D}$ indicates greater stability.  The interaction strength
parameter $k$ varies with anisotropy as $k(\lambda) = \sqrt{\lambda/0.08}$;
this ensures that the effective quasi-1D trap frequency $\omega=\lambda/4k^2$
remains equal to $0.02$ \cite{billam_etal_pra_2011}.  Rather than being
released from displaced positions in the trap, the solitary waves are launched
with a controlled velocity and relative phase using the interference method
described in Ref.\ \cite{billam_etal_pra_2011}. The computation takes advantage of the
radial symmetry of the problem, using a pseudospectral split-step method in 2D
cylindrical coordinates.}
\label{f_billam_etal_pra_2011}
\end{figure}

Again, the trapping is assumed to be cylindrically symmetric with a shape specified by the trap
anisotropy $\lambda=\omega_x/\omega_r$. As was discussed in
Section \ref{s_static3d}, the static bright solitary wave solutions in such
traps have the most soliton-like shape in the low-$\lambda$ limit
(provided the collapse threshold is not exceeded), and least soliton-like in
the opposite case. One naturally expects a similar trend in the
soliton-like-ness of the solitary wave dynamics, but is faced with the issue of
how to quantify the soliton-like-ness of the dynamics.

When considering repeated collisions between two solitary waves at the centre
of a harmonic trap, as shown in Fig.\ \ref{f_1d_collisions}, soliton-likeness
can be defined in relation to the characteristic tendency for true solitons to
emerge from mutual collisions unscathed. In Ref.\ \cite{billam_etal_pra_2011} a
metric was defined as the number of binary collisions for which the solitary
waves subsequently reach their turning point in the harmonic trap while still
having amplitudes and displacements from the origin above 75\% of their
original value. Since 1D collisions satisfy this criteria almost
indefinitely,\footnote{Eventually, the effects of the variation in the external
trap potential across the collisions could lead to the break-up of solitary
waves in the 1D GPE. However, this does not seem to occur on timescales
easily accessible to numerical simulation; instead the numerical errors grow
faster than the deviation from the soliton-like behaviour.} this metric is
termed the ``number of 1D collisions'', $C_\mathrm{1D}$.  While more
sophisticated definitions could also be employed, this simple metric for the
soliton-likeness of the dynamics is enough to reveal a rich variation in the
dynamics of collisions in the parameter space of incident velocity $v$, trap
anisotropy $\lambda$ and relative phase $\Delta\Phi$
\cite{billam_etal_pra_2011}. Fig.\ \ref{f_billam_etal_pra_2011} illustrates the
variation of $C_\mathrm{1D}$ in the $v$-$\lambda$-$\Delta\Phi$ parameter
space, for interaction strength $k = \sqrt{\lambda/0.08}$; this choice of $k$
ensures that the effective 1D trap strength used in Ref.\
\cite{billam_etal_pra_2011} is fixed to $\omega=0.02$, for which value a
solitary-wave ground state exists for $N$ atoms. Rather than being set in
motion by an initial displacement, the solitary waves in Fig.\
\ref{f_billam_etal_pra_2011} are set in motion using an interference protocol
providing arbitrary control over the relative phase and velocity (see Section
\ref{s_outlook} and Ref.\ \cite{billam_etal_pra_2011}).

As would be expected from the analysis of collisional stability in the
waveguide trap, how soliton-like the solitary wave collisions are is strongly
dependent on the relative phase $\Delta\Phi$ at low velocity, with
$C_\mathrm{1D}$ being significantly higher for $\Delta \Phi = \pi$ than for
$\Delta\Phi=0$. This phase-dependence weakens for faster collisions, which
become more soliton-like as the velocity is increased. However, the dependence
of $C_\mathrm{1D}$ on $\lambda$ is oscillatory in character. This is quite
distinct from the case when considering the \textit{static} solitary wave
ground state (Section \ref{s_static3d}), where the solitary wave shape varied
smoothly with $\lambda$.  These oscillations represent the entirely dynamical
effect of radial breathing oscillations of the solitary waves being excited
during the collision. Depending on the trap anisotropy $\lambda$ the transfer
of energy to these radial breathing oscillations can be be either enhanced or
suppressed, possibly providing an experimental ``knob'' with which to enhance
the stability of bright solitary waves
\cite{billam_etal_pra_2011}.

\subsubsection{Stability of solitary wave collisions}
\label{s_dynamic3d_trap_stability}

While the analysis of the previous section gives a comprehensive account of the
soliton-likeness of collisions for varying velocity, anisotropy and relative
phase, the interaction parameter $k$ in Fig.\ \ref{f_billam_etal_pra_2011} is
fixed (as a function of $\lambda$). Other theoretical work has explored the
$k$-dependence of solitary-wave collision dynamics, for fixed $\lambda$
\cite{parker_etal_jpb_2008,parker_etal_physicad_2008}.   It was found, for example, that even in the presence of axial trapping, the stability diagrams in $v$-$k$ space were qualitatively the same as for the axially-homogeneous case [Fig.\ \ref{f_3d_collisions_homogeneous}].  The population transfer for $\Delta \Phi \neq 0,\pi$ also occurs within an axially-trapped system, but with an additional consequence: the repetition of collisions under axial trapping leads to the continued growth of asymmetric populations between the two colliding waves, which terminates only when collapse instability occurs in one of the waves.  As such even a small deviation from $\Delta \Phi=0$ or $\pi$ can, over repeated collisions in a trap, leads to significant changes in the long term state of the system \cite{parker_etal_physicad_2008}.

These studies also performed modelling of the JILA experiment
\cite{cornish_etal_prl_2006} using the 3D GPE.  Excellent agreement with the experiment data was achieved assuming the bright solitary waves to have a relative phase very close to
$\Delta\Phi=\pi$, i.e. locally repulsive interactions.  In this manner, the solitary waves are predicted to be able to survive many collisions without collapse, as were
observed in the JILA experiment.  This is in general agreement with other work
predicting that the multiple soliton-trains seen in experiment
\cite{strecker_etal_nature_2002, cornish_etal_prl_2006} are consistent with
neighbouring solitary waves having a relative phase close to $\pi$
\cite{al_khawaja_etal_prl_2002, carr_brand_prl_2004, leung_etal_pra_2002,
adhikari_njp_2003, parker_etal_jpb_2008, parker_etal_physicad_2008}.  The origin, and even the true existence, of these $\pi$-phase differences is an open question, and will be discussed in Section \ref{s_outlook}.

\section{Hot topics in bright solitary matter waves}
\label{s_outlook}

The degree of control over the potential landscape and nonlinearity of BECs offers unique opportunities to study the fundamental properties of solitary waves, as well as nonlinear systems in general.  Furthermore, the properties of bright solitary matter waves makes them promising
candidates for a variety of future applications. Areas of current research
towards future applications include the development of soliton atom-lasers
\cite{rodas-verde_etal_prl_2005, carpentier_etal_pra_2006, chen_malomed_2005},
the stabilization and manipulation of bright solitary matter waves using
spatially and temporally varying traps and inter-atomic $s$-wave scattering
lengths \cite{saito_ueda_prl_2003, malomed_2006}, and manipulation of bright
solitary matter waves in periodic potentials \cite{poletti_etal_prl_2008,
poletti_etal_physicad_2009}, with the potential for applications in quantum
information \cite{ahufinger_etal_njp_2007}.

In this section we focus on four areas of current research: the description of
bright solitary matter waves beyond the mean-field approximation (including open questions over their relative phase), the exploitation of bright solitary matter waves in interferometry, the application of solitary waves as surface probes, and more exotic forms of bright solitary matter wave.

\subsection{Beyond-mean-field treatments and relative phase}
\label{s_manybody}

The results presented in this chapter have focussed on the mean-field, zero temperature description of a Bose-Einstein condensate provided by the Gross-Pitaevskii equation.  However, Bose-Einstein condensates are in fact many-body quantum mechanical systems at finite temperature.  To incorporate these general and important additional effects, more sophisticated models must be employed.  Such effects become particularly relevant in tightly confined geometries, e.g. a quasi-1D system, or close to the transition to Bose-Einstein condensation.  The most common family of methods to describe beyond-mean-field effects (quantum and/or thermal effects) are the various embodiments of the stochastic GPE \cite{emergent_book}.  A more fundamental quantum mechanical approach is to describe the many-body dynamics via the multiconfigurational time-dependent Hartree method for interacting bosons \cite{alon_etal_pra_2008}.

There have been limited beyond-mean-field studies of bright solitary matter waves to date, although what studies have been performed have raised intriguing differences from the mean-field predictions.

In the two experiments to date that have generated multiple bright solitary waves \cite{strecker_etal_nature_2002,cornish_etal_prl_2006}, the observed wave dynamics (post-formation of the waves) could be well described by the mean-field GPE under the assumption that the waves featured phase differences close to $\pi$ \cite{al_khawaja_etal_prl_2002,parker_etal_jpb_2008,parker_etal_physicad_2008}.   This pattern of relative phases can potentially be explained by a  process of solitary wave formation through modulational instability, followed by collapse of neighbouring solitons with relative phases close to zero
\cite{al_khawaja_etal_prl_2002, leung_etal_pra_2002, carr_brand_prl_2004,
adhikari_njp_2003}.  However, the mean-field GPE, even when supplemented with
phenomenological three-body loss terms, cannot provide
a quantitatively accurate description of solitary wave formation
out of a condensate collapse \cite{savage_etal_pra_2003, wuster_etal_pra_2005,
wuster_etal_pra_2007}.  More sophisticated simulations of the condensate quantum field during collapse recover the formation of multiple solitary waves but without $\pi$-phase differences \cite{dabrowska_wuster_etal_njp_2009}.  The origin of the $\pi$-phase differences (if they truly exist) remains an open question.

In Ref. \cite{dabrowska_wuster_etal_njp_2009} it was also observed that, under 1D quantum field simulations, that the solitary wave collisions behaved repulsively, i.e. akin to mean-field collision with $\pi$-phase difference, but independent of the initial relative phase.  Although these 1D results were not supported by the corresponding 3D quantum field simulations, uncertainty remains over whether $\pi$-phase differences do indeed appear in the experimental systems, and indeed whether relative phase as defined by the GPE is a well-defined quantity for experimental bright solitary waves.  Potentially, the relative phase and velocity dependencies predicted by the GPE could be verified in experiment
using the controlled generation method of Ref.\ \cite{billam_etal_pra_2011},
providing a test of the mean-field description.

More recently, Cederbaum {\it et al.}  used the time-dependent Hartree method  \cite{alon_etal_pra_2008} to describe the collisions between two initially-formed bright solitary waves \cite{streltsov_etal_prl_2011}.  While the GPE assumes that all atoms occupy a single quantum state, this method relaxes this condition and allows occupation of an arbitrary number of states.  It was observed that the freely evolving solitary waves rapidly lost their coherence, evolving into fragmented condensates distributed over multiple states.  These are fundamentally different objects to the original solitary waves, which within the GPE are assumed to perpetuate.  These final states can be distinguished in experiment via their signatures in the first-order correlation functions.

\subsection{Bright solitary wave interferometry}

Over the last two decades the advent of atom interferometry
\cite{cronin_etal_rmp_2009} has led to significant improvements in metrological
precision for real-world measurements of, e.g., rotation
\cite{gustavson_etal_prl_1997} and the acceleration due to gravity
\cite{peters_etal_nature_1999}. The development of atomic BECs has enabled a
new form of atom interferometry in which a \textit{trapped} BEC is coherently
split and recombined after a period of differential evolution. Following a
pioneering early experiment \cite{andrews_etal_science_1997}, many BEC
interferometers have been constructed based around the principle of a raised,
and subsequently lowered, double-well potential \cite{shin_etal_prl_2004,
schumm_etal_nature_physics_2005, wang_etal_prl_2005, fattori_etal_prl_2008,
baumgartner_etal_prl_2010}.  This scheme allows long interaction times
\cite{grond_etal_njp_2010} and the small spatial scale potentially permits
accurate measurements of, e.g., the Casimir-Polder potential of a surface
\cite{baumgartner_etal_prl_2010}.  Provided the raising of the barrier is
sufficiently fast, the GPE can provide a good description of the dynamics, in
the sense that nearly all atoms remain in a single mode, which is coherent
across the barrier \cite{faust_reinhardt_prl_2010}. However, interactions also
cause undesirable phase diffusion during the interaction time, and for this
reason experiments have typically chosen to reduce or eliminate them where
possible \cite{fattori_etal_prl_2008, grond_etal_njp_2010,
maussang_etal_prl_2010}.

The properties of bright solitary waves offer a novel solution to the problem
of inter-atomic interactions; one can envisage an analogue to the optical
Mach-Zender interferometer in which a BEC is split into two coherent,
non-dispersive, spatially-localized bright solitary waves, which are
manipulated and eventually recombined using a time-dependent external
potential. In the following sections we briefly review existing proposals for
the necessary coherent beam-splitting of solitary waves, using an
internal-state interference protocol \cite{billam_etal_pra_2011} (Section
\ref{s_split_interf}) and potential barriers \cite{weiss_castin_prl_2009,
streltsov_etal_pra_2009, helm_pra_2012, gertjerenken_etal_inprep_2011}
(Section \ref{s_split_barrier}). In Section \ref{s_surface} we outline
proposals to use interferometry devices based on bright solitary waves for
improved sensitivity in the measurement of atom-surface interactions
\cite{cornish_etal_physicad_2009}.

\subsubsection{Splitting solitary waves using interference methods}
\label{s_split_interf}

In this section we consider using a magnetic Feshbach resonance to
quasi-instantaneously change the (negative) $s$-wave scattering length from an
initial value $a_s^0$, related to the new value $a_s$ by $a_s^0 = \alpha^2
a_s$. If the BEC is initially in the bright solitary wave ground state at
scattering length $a_s^0$, it remains so immediately after the change to
scattering length $a_s$.  Assuming a 1D description and negligible axial
trapping ($\omega_x = 0$), the subsequent dynamics are described by the NLSE
[Eq.\ (\ref{e_nlse})] with initial condition,
\begin{equation}
\psi_0(x) =
\frac{1}{2\alpha\sqrt{b_x}}
\mbox{sech}\left( \frac{x}{2 \alpha^2 b_x} \right)\,,
\label{e_statmulti}
\end{equation}
where the original soliton is assumed to be centered on the origin for
convenience.  Solutions of the NLSE for this initial condition are well-known
in the context of nonlinear optics \cite{satsuma_yajima_1974}:  for integer
$\alpha=J$, Eq.\ (\ref{e_statmulti}) consists of a bound state, or multi-soliton
pulse, of $J$ solitons with unequal amplitudes $a_j$ and zero velocity ($v_j =
0$). For non-integer $\alpha =J+\beta$, where $0<\beta<1$, it consists of $J$
solitons plus radiation \cite{satsuma_yajima_1974}.

Subjecting such a pulse to a sinusoidal density modulation, such that,
\begin{equation}
\psi_0^\prime(x) =
\frac{1}{2\alpha\sqrt{b_x}}
\mbox{sech}\left( \frac{x}{2 \alpha^2 b_x} \right)
\cos \left(\frac{Kx}{2 \alpha^2 b_x} + \frac{\Delta\Phi}{2} \right)\,,
\label{e_modmulti}
\end{equation}
is the new initial condition, alters the character of the multi-soliton pulse.
For the case of most interest, $\alpha \gtrsim 2$, this modulation was explored
in Refs. \cite{kodama_hasegawa_1991, afanasjev_vysloukh_j_opt_soc_am_b_1994}
using perturbative and numerical methods. In this case, beyond a (relatively
low) threshold value of the modulation wavevector $K$ the multi-soliton pulse is split into
two solitons and a (generally negligible) radiation component. The two solitons
have equal amplitudes, oppositely directed velocities approximately proportional
to $K$ \cite{kodama_hasegawa_1991}, and relative phase $\Delta\Phi$.
Consequently such a modulation can be used to to control the velocity and
relative phase of a pair of bright solitons.

\begin{figure}
\centering
\includegraphics[width=3.8in]{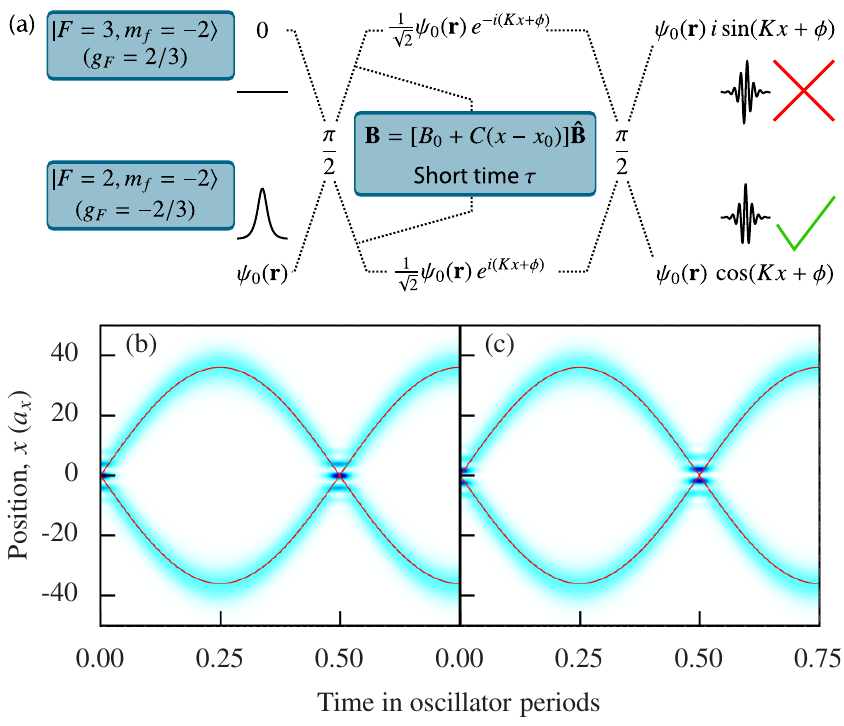}
\caption{\small Interferometric scheme to phase-coherently split a
single bright solitary wave into two bright solitary waves with controlled
velocity and phase \cite{billam_etal_pra_2011}: (a) interferometric protocol
illustrated for $^{85}$Rb and magnetic field $\mathbf{B}$, which is applied for
a short time $\tau$ between two quasi-instantaneous $\pi/2$-pulses. The
component in $|3,-2\rangle$ is not trapped, and escapes. (b) Splitting and
re-collision with relative phase $\phi=0$, and (c) $\phi=\pi$. Red lines
indicate the trajectories predicted by the particle-like model discussed in
Section \ref{s_dynamic1d} \cite{martin_etal_prl_2007}.}
\label{f_soliton_splitting}
\end{figure}

In Ref.\ \cite{billam_etal_pra_2011}, a theoretical scheme was developed to
implement such a modulation, and hence controllably generate a pair of bright
solitons, using an internal state interference protocol; this protocol is
illustrated schematically in Fig.\ \ref{f_soliton_splitting}(a). In this
protocol, the application of a linear perturbing potential for a short time is
used to impart momentum on the solitons; the waves continue to behave as
solitary waves under such a potential as shown in Ref.\
\cite{morgan_etal_pra_1997}.  Numerical simulations verified that the same
protocol is effective for bright solitary matter waves both in the presence of
an axial trap [Fig.\ \ref{f_soliton_splitting}(b,c)] and outside the quasi-1D
limit.  If such a technique can be experimentally implemented, the possibility
to carefully engineer the relative phase between two solitary waves would
represent a crucial first step towards a general bright solitary wave
interferometer.

\subsubsection{Splitting solitary waves at a potential barrier}
\label{s_split_barrier}

Compared to the interference method discussed in the previous section, a
simpler method to split bright solitary waves is afforded by collisions with a
potential barrier. In return for experimental simplicity, however, this method
lacks the same fine-grained control over the relative phase.

Within the mean field description, the dynamics of NLSE bright soliton
collisions with potential barriers and wells has been widely explored (see,
e.g., \cite{kivshar_malomed_rmp_1989, ernst_brand_pra_2010,lee_brand_2006,
cao_malomed_pla_1995, holmer_etal_cmp_2007} and Refs. therein). The behaviour
of solitary waves is similar in soliton-like regimes; in particular, fast
solitary wave collisions with a narrow barrier lead to smooth splitting of an
incoming solitary wave into transmitted and reflected solitary waves
\cite{helm_pra_2012, gertjerenken_etal_inprep_2011,
al_khawaja_stoof_njp_2011, martin_ruostekoski_njp_2012,
holmer_etal_cmp_2007}. This behaviour is analogous to bright solitons in the
NLSE scattering from a $\delta$-function potential: it can be analytically
demonstrated in such a situation that the incoming bright soliton is split into
transmitted and reflected components, each of which consist mainly of a bright
soliton, plus a small amount of radiation \cite{holmer_etal_cmp_2007}. Bright
solitary waves interacting with barriers much narrower than their width largely
follow this prediction \cite{helm_pra_2012,
martin_ruostekoski_njp_2012, al_khawaja_stoof_njp_2011}. Within the
mean-field description, potential barrier collisions of this nature have been
proposed as another means to realize solitary wave interferometers, potentially
based on solitary wave molecules \cite{al_khawaja_stoof_njp_2011}, oscillating
bright solitary waves in a harmonic trap \cite{martin_ruostekoski_njp_2012}
and bright solitary waves in a toroidal trap \cite{helm_pra_2012}.

In understanding the operation of such an interferometer one must be careful in
interpreting the predictions of the GPE. This mean-field description is most
often thought of as describing a system with a wavefunction of Hartree product
form (that is, a single macroscopically occupied single particle mode). Such a
state is free of many-body correlations \cite{streltsov_etal_pra_2009}. At face
value, this seems to be at odds with many-body descriptions of the scattering
of a bright solitary wave on a potential: it was demonstrated in Ref.\
\cite{weiss_castin_prl_2009}, using an effective potential approximation, that
a condensate bright soliton of 100 atoms could be placed in a coherent
macroscopic superposition between reflected and transmitted solitons via a slow
collision with a wide Gaussian barrier --- a state entirely dominated by
many-body correlations rather than free of them. Similar collisions were
investigated in Ref.\ \cite{streltsov_etal_pra_2009} using the MCTDHB many-body
computational method \cite{streltsov_etal_prl_2007, alon_etal_pra_2008}. In
this work, the condensate was found to fragment, leaving two macroscopically
occupied orbitals. One of these orbitals corresponded to a transmitted, and the
other to a reflected, bright soliton, implying creation of a macroscopic
coherent superposition between spatially distinct states.

This apparent disagreement with the GPE prediction is, however, only a
disagreement with the Hartree-product interpretation of the GPE; in
classical-field methods one commonly uses the GPE to describe \textit{all}
macroscopically occupied modes of a system \cite{blakie_etal_ap_2008}, and in
this interpretation there is no general disagreement with the many-body
description.  Importantly, in the proposed solitary wave interferometers
\cite{martin_ruostekoski_njp_2012, helm_pra_2012} the actual
interferometric measurement consists of measuring the average fraction of atoms
ending up on a particular side of a potential barrier. Such a measurement is
independent of the underlying occupations of single particle modes, suggesting
the mean-field GPE may still give a useful description of a bright solitary wave
interferometer.

Nonetheless, the realization of macroscopic quantum effects using bright
solitary waves offers exciting potential for future interferometric devices, as
states with macroscopic quantum superposition could be exploited to achieve
quantum enhancement of the measurement precision \cite{grond_etal_njp_2010,
maussang_etal_prl_2010}.  A bright solitary wave interferometer therefore
offers the intriguing possibility of observing the effects of macroscopic
quantum superposition \cite{dunningham_burnett_pra_2004, cooper_etal_pra_2010}
through the formation of a fragmented state \cite{weiss_castin_prl_2009,
streltsov_etal_pra_2009} and hence enhancing measurement sensitivities
\cite{dunningham_burnett_pra_2004, dunningham_contemp_phys_2006}.


\subsection{Soliton surface probes and quantum reflection}
\label{s_surface}

There is growing interest in the use of ultracold atoms to measure short range forces close to a surface motivated by the possibility of probing short range corrections to gravity which extend beyond the Standard model \cite{dimopoulos_physlettB_1996,Arkani_Hamed_physlettB_1998,Antoniadis_physlettB_1998}. Traditional experiments, following the pioneering measurements of Cavendish in 1798 \cite{Cavendish}, now use a variety of approaches from superconducting gravity gradiometers \cite{PhysRevLett.70.1195} and microcantilevers \cite{PhysRevLett.90.151101} to planar
oscillators \cite{Nature.421.922} and torsion balance experiments \cite{PhysRevLett.98.021101}. However, in scaling down experiments to probe ever decreasing length scales a new, fundamental problem arises. Quantum electrodynamics predicts a macroscopic force between conductors, known as the Casimir force \cite{Casimir.Proc.K.Ned}. This force vastly overwhelms the much weaker gravitational attraction between the test masses, such that experiments are forced to search for deviations between the theoretical and experimental Casimir forces. Precisely calculating such Casimir forces for a specific macroscopic test mass near a surface is generally difficult \cite{1367-2630-8-10-243}. In contrast, the interaction between a single neutral atom and a plane surface is well understood \cite{PhysRev.73.360,McLachlan:1963-1964:0026-8976:381} being characterised by the attractive Casimir-Polder potential,

\begin{eqnarray}
U_{\rm vdW}&=&-\frac{C_3}{z^3}  {\rm \hspace{5mm}for\hspace{5mm}} z < \lambda_{\rm opt}/2\pi, \\
U_{\rm ret}&=&-\frac{C_4}{z^4}  {\rm \hspace{5mm}for\hspace{5mm}} \lambda_{\rm opt}/2\pi < z < \lambda_{\rm T},
\end{eqnarray}
where for longer length scales the $1/z^3$ form of the van der Waals potential, characterised by $C_3$, becomes $1/z^4$ due to retardation effects. This new regime is characterised by $C_4$ with the transition point between the two regimes determined by the wavelength  corresponding to the dominant excitation energy of the interacting atoms, $\lambda_{\rm opt}$ ~\cite{CPNature}. Further from the surface (larger than the thermal wavelength of photons, $\lambda_{\rm T}$) the interaction becomes dominated by the thermal fluctuation of the electromagnetic field~\cite{PhysRevA.70.053619}.

The inherent advantage of directly probing the atom-surface interaction has prompted the recent proposal of a new generation of experiments which aim to exploit the precision and control offered by atomic physics and ultracold quantum gases to push the measurement of short-range forces into a new regime~\cite{PhysRevD.68.124021, PhysRevA.75.063608, PhysRevLett.68.3432, NJP.8.237}. Indeed a number of proof-of-principle experiments have already utilised ultracold atomic gases to explore the short range van der Waals and Casimir-Polder potentials~\cite{PhysRevLett.77.1464, PhysRevLett.70.560, PhysRevLett.104.083201, PhysRevA.72.033610}. Nevertheless such experiments are in their infancy and considerable refinement is required before they become competitive with the classical `Cavendish style' experiments as a test of short-range gravitational forces.

The attractive Casimir-Polder potential described above can also be investigated through the study of \emph{quantum reflection}. The term quantum reflection refers to the process where a particle reflects from a potential without reaching a classical turning point and is a direct consequence of the wave nature of the particle. Significant reflection occurs when the local wave vector of the particle
$k=\sqrt{(k_\infty^2-2mU(z)/\hbar^2)}$ changes by more than $k$ over a distance of $1/k$, where $k_\infty$ is the wave vector of the particle of
mass $m$ far from the potential $U(z)$. This requires an abrupt variation in the potential $U(z)$, exactly as is found for an atom in the vicinity of a solid surface. The demonstration of quantum reflection from solid surfaces is typically performed at grazing incidence in order to reduce the wave vector normal to the surface \cite{Anderson1986,Shimizu}. The advent of ultracold and quantum degenerate atomic samples with large de Broglie wavelengths opens up new possibilities to study quantum reflection at normal incidence with unprecedented control over the atomic motion. Reflection probabilities as high as $20\%$ have been demonstrated for $^{23}\mbox{Na}$ condensates incident on a solid silicon surface \cite{Pasquini}.

The use of bright matter wave solitons has been proposed to study quantum reflection from a solid surface. In Ref. \cite{cornish_etal_physicad_2009} the authors show that the use of solitons presents a number of unique advantages resulting from the presence of attractive interactions. Crucially the robust, self-trapped and highly localised nature of bright solitons can result in a clean reflection from the surface, with very limited disruption to the density profile as compared to condensates with repulsive interactions \cite{Pasquini}. Moreover, previous numerical studies of quantum reflection from purely attractive potential wells revealed that in certain regimes the whole soliton reflects with very little loss, leading to a significant enhancement of the reflection probability as compared to the single particle case \cite{lee_brand_2006}. The presence of attractive interactions has also been shown to be advantageous in the performance of traps for cold atoms based upon quantum reflection \cite{Jurisch}. The absence of dispersion as the soliton propagates permits the precise control of the velocity normal to the surface and allows much lower velocities to be achieved.

\begin{figure}
	\centering
		\includegraphics[width=\textwidth]{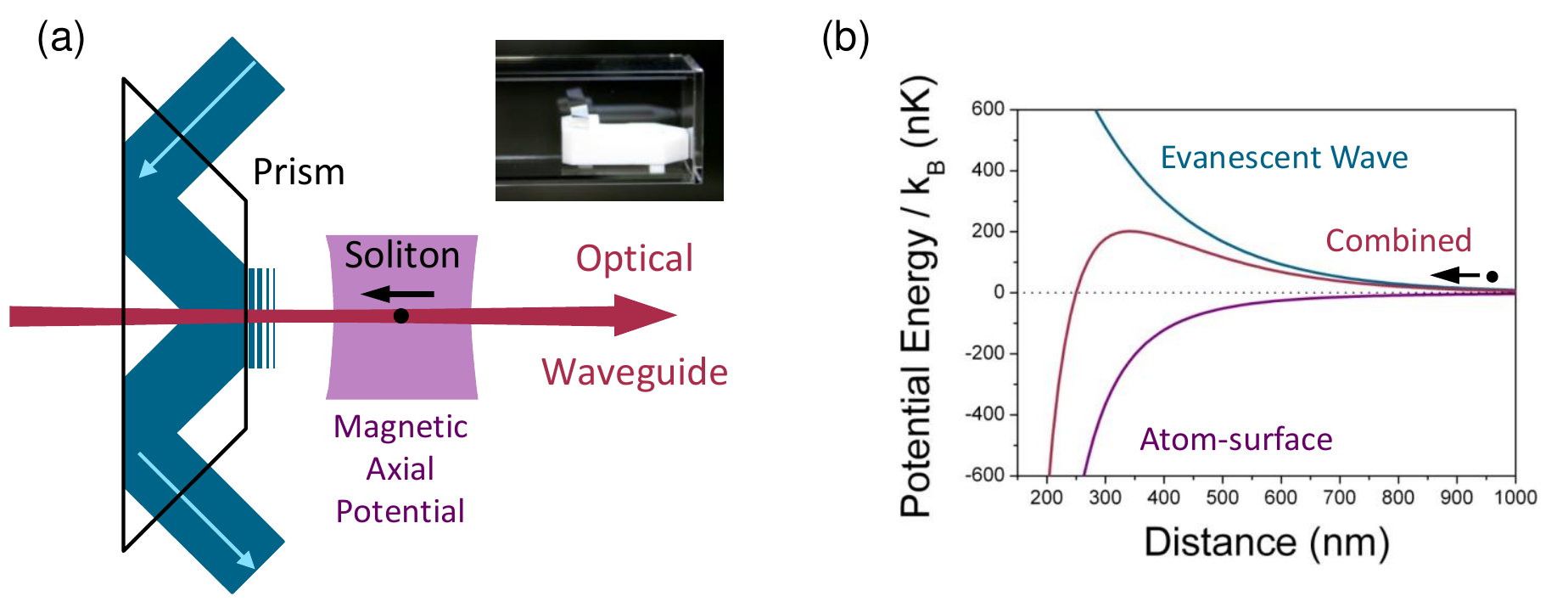}
	\caption{\small (a) Schematic of the proposed experimental configuration for the study of quantum reflection of bright matter wave solitons from a solid surface. The soliton
propagates towards the surface in an optical waveguide formed by a focussed $1064\,\mbox{nm}$ laser beam. Motion of the soliton along the waveguide is controlled through the manipulation of a weak magnetic potential along the waveguide. An optional repulsive evanescent field can be added through the total internal reflection of a $532\,\mbox{nm}$ laser beam in the prism. The inset shows a photograph of the super-polished Dove prism mounted in a UHV glass cell in the Durham experiment.  (b) The total potential (red) experienced by the atoms in the vicinity of the surface is the sum of the Casimir-Polder potential (purple) and the evanescent field (blue).}
	\label{fig:surface}
\end{figure}

Current experiments in Durham \cite{Durhamapparatus,GuidedTransport} aim to exploit this combination of advantages and promise to deliver accurate measurements of the quantum reflection probability.  The proposed experimental scenario for the study of quantum reflection is depicted in Fig.~\ref{fig:surface}.

Bright matter wave solitons are formed in an optical waveguide from an $^{85}$Rb condensate with attractive interactions.For rubidium and a room temperature surface, the lengthscales relating to the Casimir-Polder potential are $\lambda_{\rm opt}/2\pi\approx0.12\,\mu$m and $\lambda_{\rm T}\approx 7.6\,\mu$m.  The velocity of the soliton towards the surface can be controlled by manipulating a weak ($\sim1$\,Hz) magnetic potential along the axis of the waveguide. The experimental configuration also allows for the addition of a repulsive (or attractive) evanescent field in the vicinity of the surface formed by the total internal reflection of a blue (or red) detuned laser field within the glass prism. This produces a potential which decays exponentially with distance from the surface; the decay length being determined by the laser wavelength, the refractive index of the prism and the angle of incidence of the laser beam. When combined with the atom-surface potential, the repulsive evanescent field leads to a repulsive barrier of finite height, in close proximity to the surface (see Fig.~\ref{fig:surface} (b)). Studies of classical reflection from such a barrier can be used to probe the atom-surface potential \cite{PhysRevLett.77.1464}. Moreover, the addition of both repulsive and attractive evanescent fields can be used to engineer a potential in the vicinity of the surface that significantly enhances the quantum reflection probability \cite{PhysRevA.67.041604}.

\subsection{Exotic bright solitary waves}
\label{s_exotic}

We have focussed in this chapter on bright solitary waves of a single-species condensate with {\it s}-wave interactions.  However, part of the beauty of these atomic gases is that it is possible to precisely introduce additional complexity into the system, e.g. additional condensate components or  long-range interactions.    In certain such scenarios, distinct bright soliton-like structures can arise, with even richer properties than their {\it s}-wave counterparts.  We will briefly summarise these exotic bright solitary waves below.

\paragraph{{\it Bright-dark solitons}}
Condensate systems which involve two co-existing condensates, composed of either different atomic species or the same atomic species but in two different magnetic states, have been the topic of much experimental and theoretical study \cite{emergent_book}.  As well as the local intra-species interactions within each component, there exists a local inter-species interaction.  When all interactions are repulsive, and the inter-species interaction is sufficiently strong, phase separation of the condensates becomes favourable \cite{pethick_smith_2002}.  Then, the system can enter a state where one component adopts a bright soliton-like structure, about which the other component forms a dark soliton-like density notch.  These hybrid structures, termed dark-bright solitons, have analogs in nonlinear optics \cite{kivshar_1998} and were first predicted to exist in inhomogeneous BECs by Busch and Anglin \cite{busch_prl_2001}.  Since then they have been generated and observed in several experiments \cite{anderson_prl_2001,becker_nature_physics_2008,hamner_prl_2011}.  Possessing only repulsive interactions, they are free from the collapse instability and behave more akin to dark solitary waves (see reference \cite{frantzeskakis_2010} for a review) than bright solitary waves.

\paragraph{{\it Bose-Fermi solitons}}
Similar to above, it is possible to create a system in which a Bose-Einstein condensate co-exists with a degenerate gas of fermions.  For identical fermions, the Pauli exclusion principle prevents {\it s}-wave interactions, and so the predominant interactions in this ultracold mixture are the {\it s}-wave boson-boson interactions and the boson-fermion interactions.  Here, an attractive boson-fermion interaction can support a soliton structure in which localized wavepacket of each gas, overlapping in space, is self-trapped by their mutual interaction.  These structures, termed Bose-Fermi solitons, have been predicted via numerical and variational approaches \cite{karpiuk_prl_2004,adhikari_pra_2005,adhikari_pra_2007,salasnich_pra_2007,parker_pra_2012}, and have been simulated to propagate without dispersion \cite{karpiuk_prl_2004,karpiuk_pra_2006}.   The boson-fermion interaction must be sufficiently strong to overcome the internal repulsion within the BEC component but not so large as to induce collapse.  As such, in 3D, they exist as metastable states of the system.  Interestingly, under the inclusion of higher order {\it p}-wave interactions with repulsive sign, the collapse instability can be completely eradicated, suggesting a greatly enhanced stability \cite{parker_pra_2012}.  Related bright solitary wave structures are also predicted to arise in Bose-Fermi mixtures in the presence of optical lattice potential \cite{salerno_pra_2005,bludov_pra_2006,adhikari_pra_2007b}.

\paragraph{{\it Non-local bright solitons}}

In recent years, BECs have been produced in which the atomic species have a large natural magnetic dipole moment \cite{lahaye_2009}.  In contrast to {\it s}-wave interactions which are short-range and isotropic, these interactions are long-range and anisotropic. By polarizing the dipoles in a common direction, the whole condensate takes on a dipolar nature, which can be accounted for within the GPE by the inclusion of a non-local term \cite{lahaye_2009}.  The attractive component of the dipolar interactions lends itself to support self-trapped solitary wave states, but it also makes the collapse instability a general feature of dipolar BECs.

Within a quasi-1D waveguide, solitary waves of dipolar BECs are predicted to be supported \cite{cuevas_pra_2009,young_2011}.  In general, dipolar interactions also co-exist with {\it s}-wave interactions, and the capacity for self-trapping was shown to occur for various regimes of the dipolar interaction and the {\it s}-wave interaction.  Collisions between two such dipolar solitary waves were found to exhibit more complicated dynamics including regimes where the colliding waves form a bound state \cite{cuevas_pra_2009} and sensitivity to the polarization direction \cite{young_2011}.  More strikingly, in Refs. \cite{pedri_prl_2005,tikhonenkov_prl_2008} it was shown that a dipolar BEC within a quasi-two-dimensional trap can form a bright solitary wave that is self-trapped in two-dimensions and free to move within the untrapped plane of the system.

Distinct non-local effects can be introduced by coupling of a ground state BEC to highly excited Rydberg states \cite{pupillo_prl_2010}, excited atomic states with high principal quantum number.  These excited states become coherently shared throughout the condensate, inducing a strong collective interaction via van der Waals forces.  Using a GPE which incorporated an appropriate non-local term, it was predicted in \cite{maucher_etal_prl_2011} that bright solitary waves could form which are not only free from the collapse instability, but are self-trapped in all three dimensions. These wavepackets are predicted to remain stable for hundreds of milliseconds and raise the prospect of the first realization of 3D bright solitary matter waves.

\section{Conclusions}
\label{s_conclusions}

We have reviewed bright solitary waves composed of gaseous Bose-Einstein condensates, from their experimental formation and observation, through to a theoretical exposition of their static and dynamical mean-field properties.   Emphasis is placed on how the harmonic trapping potential and three-dimensional setting leads to departures in behaviour from the classic bright soliton.  Soliton-like states remain supported, in the sense that they are capable of self-trapping and retaining their static form as they propagate.  The most marked deviation is introduced by the extension from 1D to 3D, which introduces an instability to collapse for sufficiently large interaction parameter and a population transfer during collisions.  These deviant behaviours can be greatly reduced in appropriate regimes but may be exploited in their own right, e.g. by using population transfer as a means to infer relative phase.

The properties of these bright solitary matter waves are now, generally speaking, well understood at the mean-field level.  Emphasis is now turning to developing a full quantum mechanical understanding of these excitations and much work remains to be done in this direction.  The experimental capacity to engineer additional atomic components and interactions into the system promises new families of bright solitary waves, which are, for instance, self-trapped in higher dimensions and stable to collapse.  Furthermore, we believe that bright solitary waves hold strong potential as atomic vehicles for applications in matter wave interferometry and surface force detection.

\bibliographystyle{plain}

\end{document}